\documentclass[11pt,epsf]{article}

\usepackage{graphicx}

\usepackage{amsbsy}

\def\beq{\begin{eqnarray}}
\def\eeq{\end{eqnarray}}
\def\bea{\begin{eqnarray*}}
\def\eea{\end{eqnarray*}}

\def\centeron#1#2{{\setbox0=\hbox{#1}\setbox1=\hbox{#2}\ifdim
\wd1>\wd0\kern.5\wd1\kern-.5\wd0\fi
\copy0\kern-.5\wd0\kern-.5\wd1\copy1\ifdim\wd0>\wd1
\kern.5\wd0\kern-.5\wd1\fi}}
\def\ltap{\;\centeron{\raise.35ex\hbox{$<$}}{\lower.65ex\hbox{$\sim$}}\;}
\def\gtap{\;\centeron{\raise.35ex\hbox{$>$}}{\lower.65ex\hbox{$\sim$}}\;}
\def\gsim{\mathrel{\gtap}}
\def\lsim{\mathrel{\ltap}}

\def\singleandthirdspaced{\baselineskip=\normalbaselineskip\multiply
    \baselineskip by 130\divide\baselineskip by 100}

\newcommand{\mymatrix}[1]{\left(\begin{matrix}#1\end{matrix}\right)}
\newcommand{\mypmatrix}[1]{\begin{pmatrix}#1\end{pmatrix}}
\newcommand{\f}{f_{ij}}
\newcommand{\p}{{\cal P}}
\newcommand{\F}{{\cal F}}

\newcommand{\e}{e}

\usepackage{amsmath}
\usepackage{amsfonts}
\usepackage{amsbsy}
\usepackage{eufrak}

\begin{document}
\begin{titlepage}
\begin{flushright}
{\large hep-ph/0505222 \\ WIS/12/05-MAY-DPP, LBNL-57533\\
}
\end{flushright}

\vskip 1.2cm

\begin{center}

  {\LARGE{\bf Split Fermions Baryogenesis from the Kobayashi-Maskawa Phase}}

\vskip 1.4cm

{\large  Gilad Perez$^*$ and Tomer Volansky$^\dagger$}
\\
\vskip 0.4cm
{\it $^*$Theoretical Physics Group, Ernest Orlando Lawrence Berkeley
  National Laboratory, Berkeley, University of California, CA 94720} \\
{\it $^\dagger$Weizmann Institute of Science \\ Rehovot, Israel}
\\

\vskip 4pt

\vskip 0.4cm

\begin{abstract}
  A new scenario of baryogenesis is presented, within the split
  fermions framework. Our model employs a first order phase transition
  of the localizer field.  The standard model (SM), Kobayashi-Maskawa
  phase induces a sizable CP asymmetry. The usual suppression of CP
  violation which arises in the SM baryogenesis is absent due to the
  existence of order one Yukawa couplings before the fermions are
  localized in the extra dimension.  Models of the above type
  naturally contain B-L violating operators, allowed by the SM
  symmetries, which induce the baryon asymmetry.  Our mechanism
  demonstrates the following concept: the flavor puzzle and the SM
  failure to create the baryon asymmetry are linked and may have a
  common resolution which does not rely on introduction of new CP
  violating sources.


\end{abstract}

\end{center}

\vskip 1.0 cm

\end{titlepage}
\setcounter{footnote}{0} \setcounter{page}{2}
\setcounter{section}{0} \setcounter{subsection}{0}
\setcounter{subsubsection}{0}

\singleandthirdspaced

\section{Introduction}
\label{sec:introduction}
It was understood long ago that the SM contains all the
ingredients~\cite{Sakharov:1967dj} required to produce the baryon
asymmetry of the universe (BAU)~\cite{Eidelman:2004wy} \beq {n_{\rm
    B}\over s}\sim 7\times 10^{-11}\,, \eeq where $n_{\rm B}$ and $s$
are the baryon and entropy densities respectively.  Nevertheless the
SM fails to explain the observed BAU quantitatively.  The lower bound
on the Higgs mass~\cite{Lep2,Eidelman:2004wy} implies that the EW
phase transition (EWPT) is second order~(see e.g.
\cite{Kajantie:1996mn,GSW} and refs. therein).  Thus no departure from
thermal equilibrium is obtained. Furthermore, the observed quark
flavor parameters are small and hierarchical, posing a puzzle, known
as the SM flavor puzzle. A measure of their smallness can be given by
the Jarlskog determinant~\cite{Jarlskog:1985ht}, ${\cal J}=\det[Y_u
Y_u^\dagger,Y_d Y_d^\dagger]={\cal O}\left(10^{-19}\right)$.  In the
SM baryogenesis case, the amount of CP violation (CPV) produced is
suppressed by ${\cal
  J}$~\cite{Nelson:1991ab,Farrar:1993sp,Farrar:1993hn,Gavela:1993ts,Huet:1994jb}.
Thus even under the assumption of a first order phase transition
(PT), the possibility that the SM can account for the
observed BAU is precluded.

It is interesting that the failure of SM baryogenesis is directly
related to the SM flavor puzzle.  Recently it was demonstrated
in~\cite{Berkooz:2004kx} that a class of
Froggatt-Nielsen~\cite{Froggatt:1978nt} models which solve the SM
flavor puzzle can also account for the BAU.  The main idea
of~\cite{Berkooz:2004kx} is that within the Froggatt-Nielsen framework,
Yukawa couplings can vary with temperature. Thus it is possible that
in the early universe the flavor parameters were anarchical, ${\cal
  J}\sim 1$. This implies that CPV arising from the Kobayashi-Maskawa (KM)
phase, during the electroweak phase transition, is unsuppressed.

In this work we demonstrate that the five dimensional (5D) split
fermions framework can realize the above idea in a similar
manner.
In
our scenario, fermions are localized due to their couplings to an
$x_5$ dependent VEV of a bulk scalar denoted as the
localizer~\cite{Georgi:2000wb,Perez:2002wb,Grossman:2002pb}.  We assume that
at the critical temperature, two phases coexist where in one of
these phases, the localizer's VEV is zero (the symmetric phase) while in the other
it acquires a non-trivial VEV (the broken phase).  As in the standard
EW baryogenesis, the phase transition occurs through bubble
nucleation.  Outside the expanding bubble wall
(in the unbroken phase) the fermions wave functions are
flat~\cite{Masiero:2000fr,Chung:2001tp,NP1} and the Yukawa couplings are of order one.
The KM phase may therefore be sufficient to explain the observed
baryon asymmetry.

Flavor models typically have two distinct phase transitions. The first
is related to the scale at which the scalars, which are SM singlets and which control
the hierarchy in the flavor sector, acquire VEVs. In our model, we
denote this phase transition as the localizer phase transition (LPT).
The other is the celebrated EWPT.  In that context, there are two
interesting related possibilities:
\begin{itemize}
\item[(i)] The LPT occurs at or below the temperature of the
  EWPT.  
\item[(ii)] The LPT occurs at temperatures well above the temperature
  of the EWPT~\cite{NP1}.
\end{itemize}
Case (i) is very similar to the one discussed
in~\cite{Berkooz:2004kx}.  Case (ii) differs from the SM baryogenesis
scenario. We show below that it can also naturally avoid all the
failures of the SM baryogenesis mechanism and account for the observed
BAU.  In our framework, we do not require new sources of CPV or new
sources which violate baryon number.  The related sources within the
SM are shown to be sufficient.  Furthermore, generically, we expect
that in simple split fermions models, option (ii) would be realized as
follows: naturally the critical temperature for the LPT, $T_{\rm
  LPT}$, will be set by the fundamental scales of the problem ${\cal
  O}(1/\pi R)\lsim T_{\rm LPT}\lsim M_*$ (where $R$ is the radius the
extra dimension).  Phenomenologically, the inverse radius, $1/R$ is
constrained to be rather high, say above $100\,$TeV, in order to
suppress the contributions to various flavor changing
processes~\cite{FCNC,KT}.  We then expect that $T_{\rm LPT}$ would be
of similar order and therefore much above the EW symmetry breaking
scale.  This implies that the BAU is produced during the LPT, well
before the EWPT. Let us briefly describe how Sakharov's conditions are
fulfilled in our scenario: C and CP violation is of order one in the
unbroken phase since all the Yukawa couplings are of order unity.  Out
of equilibrium is obtained assuming a first order PT.  This can be
accounted for by, for example, the presence of higher dimension
operators in the localizer's potential~\cite{NP1}.  Baryon violation
occurs since there are non-renormalizable operators which are not
suppressed any further by the wave-functions overlapping.  It is
remarkable that our mechanism goes through even in the most minimal
split fermions models a la Kaplan-Tait~\cite{KT}. We shall see that,
despite the fact that in these models the localizer's couplings are CP
conserving~\cite{NP1,HPSS,GHPSS}, a sizable baryon asymmetry is
obtained.

In section~\ref{sec:framework} we introduce the split fermions
framework and in particular we discuss in some detail the specific
model we study to demonstrate our mechanism.  In
section~\ref{sec:cosmological-setup} we describe our mechanism and
explain how we derive a semi-quantitative estimation for the resultant
asymmetry.  In section~\ref{sec:bary-numb-viol} we show how baryon
number violation occurs in our model. This is due to higher
dimensional B$-$L violating operators allowed by the SM symmetries.
In section~\ref{sec:cp-asymmetry} we present two complimentary methods
to estimate the CP asymmetry due to the interaction with the bubble
wall.  Many of the detailed calculations relevant to this section are
found in the appendices.  We conclude in
section~\ref{sec:conclusions}.

\section{The Framework}
\label{sec:framework}
In this part we briefly review some of the relevant
features of the split fermions framework.
To demonstrate our mechanism we
consider a minimal model with a single localizer scalar.
To first order the localizer's
profile is roughly flat~\cite{Georgi:2000wb,Perez:2002wb} so that at zero temperature
such a model is similar to the one in ref~\cite{KT} with constant odd masses as
described below.

We consider one extra dimension compactified on $S^1/{\mathbb Z}_2$
with a fundamental domain $[0,\pi R]$.
The matter content is that of the SM with an additional real scalar
``localizer'' field $\phi$.  The relevant part of the 5-dimensional Lagrangian density is
given by
\begin{equation}
  \label{eq:37}
  {\cal L}_5 = \bar\Psi_i\left[i\delta_{ij}\gamma^M\partial_M -
  \frac{f_{ij}}{M_*^{1/2}}\phi\right]\Psi_j +
  \frac{1}{2}\partial_M\phi\partial^M\phi-V(\phi),
\end{equation}
where $M = 0,..,3,5$,  $\Psi_i$ denote SM four-Dirac fermions (which more explicitly
are $Q_i, u_i, d_i, L_i, e_i$ with corresponding couplings $f^Q_{ij},
f^u_{ij}$ etc., where here and below the representation index is suppressed), $i,j = 1,..,3$ are
the flavor indices, $M_* \sim 10/\pi R$ is the fundamental scale
in the theory and we take $\gamma^5 =
-\gamma^0\gamma^1\gamma^2\gamma^3$. $V(\phi)$ is the localizer's
potential and a $\phi^6$ term is required to obtain a first order
PT~\cite{NP1} 
(see also \cite{GSW} for a recent study of the 4D case).
Due to the 5D Lorentz symmetry the $f$s are hermitian
and therefore can all be brought to a basis in which they are
real and diagonal.  Thus CP is not broken by the above action.
In more complicated (and realistic) models~\cite{Grossman:2002pb} twisting
of the fermion wave functions might occur in flavor
space~\cite{GHPSS}, providing new CPV sources~\cite{NP1,HPSS}.
We shall see below, that even in the limit where these
additional sources are switched off, the above framework can
still account for the observed BAU.

To avoid the constraints from FCNC~\cite{FCNC,KT} $1/R\gsim 100\,{\rm
  TeV}$ is required.  Note, however, that no upper bound on either
$1/R$ or $M_*$ exist.  For naturalness we assume that all the
dimensionless couplings in the theory are of order one.  We emphasize
that all our computations below are done at tree level and we assume
that quantum effects are sub-dominant.
  
The orbifold boundary
conditions are
\begin{eqnarray}
  \label{eq:38}
  \phi(x^\mu, -x^5) = -\phi(x^\mu, x^5),&& \;\;\; \phi(x^\mu ,x^5+ \pi R) =
  -\phi(x^\mu, x^5-\pi R),
  \nonumber \\
  \Psi(x^\mu, -x^5) = \pm \gamma_5\Psi(x^\mu, x^5),&&\;\;\; \Psi(x^\mu, x^5+\pi R) =
  \pm\gamma_5\Psi(x^\mu,x^5-\pi R), 
\end{eqnarray}
where the $+(-)$ signs are for $SU(2)$ doublets(singlets).  These
boundary conditions are intended for two things.  For the fermions,
these conditions project one of the two chiralities of the zero mode,
rendering a chiral low-energy theory.  For the localizer the
conditions impose a non-trivial $x^5$-dependent VEV, which to our
purpose is assumed to be a step function.  Thus it 
dynamically plays the role of the odd mass term, $\alpha_i$ of
ref~\cite{KT} (for more precise analysis see {\it e.g.}
\cite{Georgi:2000wb,Perez:2002wb,Grzadkowski:2004mg}).  As a consequence of this non-trivial VEV for
$\phi$, the fermions zero-modes are localized in the extra dimension,
\begin{equation}
  \label{eq:71}
  \Psi^{(0)}_i(x^5) =
  \sqrt{\frac{2\alpha_i}{1-e^{2\pi\alpha_i R}}}\left\{\begin{array}{lllll}
      e^{-\alpha_i x^5} &&&& f_{ii} > 0 \\ e^{-\alpha_i (\pi R - x^5)} &&&&
  f_{ii} < 0
  \end{array}\right.
\end{equation}
where $\alpha_i = |f_{ii}u|/M_*^{1/2}$, and $f_{ii}$ is the eigenvalue of
the Yukawa matrix $f_{ij}$ which corresponds to the eigenvector
$\Psi^{(0)}_i$ (note that $i$ should not be summed in the above). The
small overlap between the different zero modes generates the flavor
hierarchies.  In the model we consider, the hierarchies are realized
by taking for the weak doublet fields $f_{ii} > 0$ and for the weak
singlets $f_{ii} < 0$~\cite{KT}.

\section{Cosmological Setup}
\label{sec:cosmological-setup}
Below we describe the baryon-asymmetry production mechanism and
explain how the above framework 
overcomes the SM difficulties.
Our main point here is that, even minimal, split fermions models
can efficiently produce baryon number through the LPT.
It is the fact that during the LPT, near the expanding localizer
bubble wall, two different phases coexist which essentially allows for
all of Sakharov's conditions to be realized and lead to 
a sizable baryon production rate. In the unbroken phase
CPV and B$-$L production rates are unsuppressed while
inside the bubble the relevant processes freeze out due to fermion localization.

It is important to identify the time, in the history of the Universe,
in which the baryon production occurs.  This is directly linked with
$T_{\rm LPT}$, the critical temperature of the LPT.  As mentioned
above we naturally expect ${\cal O}(1/\pi R)\lsim T_{\rm LPT}\lsim
M_*$.  This is also supported by the phenomenological requirement of
the model as follows.  The flavor hierarchy is obtained due to the
geometrical setup of the fermions in the extra dimension.  To account
for that, the localizer's mass and other fundamental parameters in its
potential,
should be of order ${\cal O}(10/\pi R)$ (to have strong enough
localization~\cite{Perez:2002wb,Grossman:2002pb}).  In this case we
then expect that the critical temperature $T_{\rm LPT}$ would be of
similar order.  To avoid constraints from processes which induce
flavor changing neutral currents the inverse radius is constrained to
be rather high, $1/R\gsim 100\,$TeV~\cite{FCNC,KT}.  and therefore the
LPT occur well before the EWPT, \beq T_{\rm LPT}\gg T_{\rm EWPT}\,.
\eeq We shall assume below that $T_{\rm LPT}\sim 1/\pi R$. This
implies that during the LPT, modes with very high 5D momentum are
Boltzmann suppressed. Hence the dynamics of the baryon production can
be roughly estimated using our knowledge of 4D field theory.

Let us discuss in some more detail how Sakharov's conditions
are satisfied within our framework.
\begin{itemize}
\item {\emph{ CP violation} - } Outside the bubble wall the quarks
  wave functions are flat and thus the corresponding Yukawa matrices
  are naturally sizable and anarchical. The presence of these sizable,
  flavor breaking, couplings modify the quark thermal
  masses~\cite{Weldon:1982aq,Farrar:1993sp,Farrar:1993hn,Huet:1994jb}
  and imply that CPV is unsuppressed.  Since however the LPT occurs
  before the EWPT it is the fact that the quarks have non-universal, CP
  conserving, couplings to the localizer which drives the CP
  asymmetries. This is discussed in more details below.
\item {\emph{ B violation} - } Above the EWPT any B+L asymmetry
  created is washed out by sphalerons.  It is therefore necessary for
  B$-$L violating interactions to exist and contribute to the
  asymmetry.  It is remarkable that no new ingredients are required to
  induce B$-$L number.  In fact it was noted a while ago, by
  Weinberg~\cite{Wein}, that non-renormalizable B$-$L violating terms
  exist in the SM.  These interactions are only mildly suppressed outside the
  bubble and are frozen in the bubble after fermions localization has
  occurred.  In fact, the B$-$L operators in our case play the same
  role which is played by the sphaleron processes in the SM
  baryogenesis. Consequently, we expect that they would yield baryon
  production in our case.
\item {\emph{Deviation from thermal equilibrium} - } First order phase
  transition can be accounted for by the presence of a $\phi^6$ term
  in the 5D effective localizer's potential~\cite{NP1}. This case was
  recently studied in great details for the SM 4D case~\cite{GSW}
  where it was indeed found that such terms induce a first order phase
  transition and consequently deviation from thermal equilibrium.
\end{itemize}

Let us now briefly describe how the mechanism of baryon production goes
through in our model.
At $T_{\rm LPT}$ a phase transition occurs
and a bubble of the true vacuum is expanding through space.  Just like
in the SM baryogenesis, as the wall sweeps through space, particles
and anti-particles hit the bubble wall, interacting with the localizer
field.  From the four dimensional point of view, there is a KK tower
of particles and as we show in section \ref{sec:cp-asymmetry}, the
interactions with the wall violate KK number conservation, so an
even(odd) incoming KK mode is reflected into an odd(even) KK mode.
CP violation implies that the reflection amplitudes for particles will
be different than the ones for
anti-particles resulting in an asymmetry in the corresponding
densities near the interface. 
This asymmetry induces density of a conserved global charge
which decays far from the bubble wall due to decoherence effects~\cite{HT}.
The induced charge density, in the unbroken phase, biases the B$-$L violating
interactions towards production of B$-$L net
density~\cite{HT,Nelson:1991ab}.
This B$-$L production is shut-off when the expanding bubble wall
quickly overtakes the region of non-zero charge and
the net B$-$L charge eventually becomes
our observed BAU.

An estimate for $n_B/s$ may be derived in a similar manner to the one
presented in~\cite{Nelson:1991ab,Huet:1994jb}.  For simplicity, we assume a $2+1$
dimensional problem with $(x^0, x^3, x^5)$ spanning space.  Thus the
computation of the reflection and transmission of particles in
the non-compact space becomes a one dimensional problem.  Furthermore, we
work in the thin wall approximation with low wall velocity,
$v_{\rm w}\simeq 0.1$. 
Given the above assumptions one obtains~\cite{Huet:1994jb},
\begin{equation}
  \label{eq:42}
  \frac{n_B}{s} \simeq
  -\epsilon_{\text{B-L}}\left(\frac{15}{2\pi^2g_*}\right)\sum_{n,m}\int\frac{dE}{2\pi
    T}\,\left[(n_{n+}^u - n_{m-}^u) - (n_{n+}^b - n_{m-}^b)\right]
  \times \Delta_{\rm CP}^{nm}(E),
\end{equation}
where $\epsilon_{\text{B-L}}$ is a suppression arising from the
inefficiency of the B-L violating interactions near the expanding bubble wall
and $g_*\sim 10^2$ is the
number of relativistic degrees of freedom.  $n_{n\pm}^{u(b)}$ is the
Fermi-Dirac distribution boosted to the wall's rest frame, $n_{n\pm} =
\left[e^{(E-v_{\rm w}p_3^{(n)})/T}+1\right]^{-1}$ for a n-KK mode of helicity
$\pm$ in the symmetric(broken) phase.  Perturbatively, for a small
Yukawa coupling to the localizer one can estimate,
\begin{equation}
  \label{eq:45}
  \left[(n_{n+}^u - n_{m-}^u) - (n_{n+}^b - n_{m-}^b)\right] \simeq 
  2n_0(E)\left[1-n_0(E)\right]\,\left[p_3^{(n)}-p_3^{(m)}\right]\,\frac{v_{\rm w}}{T}\,,
\end{equation}
with $p_3^{(n)}$ the momentum in the $x^3$ direction of the n'th KK
mode and $n_0(E) = (e^{E/T}+1)^{-1}$.  This is not true in
the large coupling limit due to large corrections to the fermion
masses, and hence to the momentum.  Finally,
\begin{equation}
  \label{eq:43}
  \Delta_{\rm CP}^{nm}(E) = {\rm Tr}\left[R_{nm}^\dagger R_{nm}^\dagger -
  \bar R_{nm}^\dagger\bar R_{nm}\right]
\end{equation}
is the CP asymmetry related to the reflection coefficients.  The
quantity $R_{nm}$ is a matrix in flavor space that contains the
reflection coefficients.  For example, $R_{nm}^{if}$ is the
coefficient for a reflection of an $n$th KK quark state (with, say,
helicity $+$) and flavor $i$ into an $m$th KK quark state (with
helicity $-$) and final flavor $f$.  $\bar R_{nm}$ corresponds to the
$CP$-conjugate processes.  In deriving the equation above, we have
used the unitarity relation $\sum_m (R_{nm}^\dagger R_{nm} +
T_{nm}^\dagger T_{nm}) = $1 where $T_{nm}$ is the transmission
coefficient for the nth KK mode.  As we show in Appendix \ref{apUnit}
this relation obtains corrections of order $p_3^{(n)}/T$ which are of
order unity.  Thus we expect order one corrections to our
estimation of $n_B/s$.  In section \ref{Nonpert} and appendices
\ref{sec:non-pert-calc-1} and \ref{apUnit}.  we present a
non-perturbative analysis which supports our claim that the above
corrections are under control.

A remark is now in order.  The CP asymmetry has contributions from
scattering of either the quark doublets or the quark singlets.  Each
of the three fields ($Q, u, d$) has an independent coupling to the
localizer, and different Yukawa interactions which produce slightly
different asymmetries.  For simplicity, we calculate below the
asymmetry for the quark doublets, however, Yukawa interactions mix
between the doublets and singlets and since these interactions are in
equilibrium during baryogenesis, the total CP asymmetry is roughly
given by the (weighted) average between the CP asymmetries of both doublets and singlets.

\section{Baryon Number Violation}
\label{sec:bary-numb-viol}
As we explained in section \ref{sec:cosmological-setup}, a successful
baryogenesis scenario requires a source of B$-$L violation.  Indeed,
since the critical temperature $T_{\rm LPT}$ is well above the EW
symmetry breaking scale, sphaleron processes are fast both inside and
outside the bubble.  Consequently, the only way to produce baryon
number which will not be washed out by sphaleron processes is via B$-$L
violating interactions.  Furthermore the B$-$L violating processes which
are fast in the symmetric phase must be very slow in the broken phase
in order to produce the asymmetry.  The split
fermions framework incorporates such B$-$L violating processes through
non-renormalizable operators.  Below we systematically go through the
list of these operators, in order to determine which of these
is important for our mechanism.

Starting from the lowest dimensional operator, dimension seven in 5D,
one finds the celebrated $L^2 H^2$.  This operator induces Majorana
neutrino masses, but is not directly relevant to baryon production.
Such an operator, however, may not only generate too heavy neutrino
masses but also washout excess of B$-$L number and therefore must be
suppressed (at least in the broken phase) for our scenario to work.
This can be realized using additional extra dimension, a discrete
symmetry~(see {\it e.g} \cite{Dis} for a recent discussion) or a
continuous one~\cite{NP1} giving the lepton doublets some charge so
that this term is highly suppressed. Another alternative is if the
Higgs is localized away from the lepton doublet.  Whatever the
mechanism is, we assume that it does not interfere with the
baryon production mechanism discussed below.

It was noticed long ago that the symmetries of the SM allow for B$-$L violating
operators~\cite{Wein} of
dimension 9 and 9$1\over2$ in 5D.
For completeness, we list these operators
\begin{eqnarray}
  \label{eq:12}
  \text{Dimension 9:} && (\overline{d^c} D_M d)(\overline L \gamma^M Q)  \,, \ \
  \ (\overline L\,D_M\,d)(\overline{d^c} \gamma^M Q) \,, \ \ \ (\overline{d^c}
  \,D_M\,d)(\overline e\gamma^Md) 
  \\
  \label{eq:15}
  \text{Dimension 9$\frac{1}{2}$:} &&(\overline{Q^c}Q)(\overline L d)
  H^\dagger\,, \ \ \ (\overline{d^c}d)(\overline e Q)  H^\dagger \,, \
  \ \ (\overline{d^c}d)(\overline L u)  H^\dagger
  \nonumber \\ 
  &&(\overline{d^c}u)(\overline L d)  H^\dagger  \,, \ \ \ (\overline{d^c}d)(\overline L d) H 
\end{eqnarray}
where $Q$($L$) stands for a quark(Lepton) doublet, $u,d$$(e)$
stand for the up and down quark(lepton) singlets, $H$ stands for the Higgs
field, $D$ stands for a covariant derivative, c denotes charge conjugation and $M=0..3,5$ stands for
Lorentz indices. In the above 
we have suppressed color, flavor and  $SU(2)$ indices.

Let us estimate the rate for the dimension 9 B$-$L violating operators
of eq. \eqref{eq:12}.  Using naive dimensional analysis (NDA) we
estimate the coefficient of the above operators as $(4\pi)^2$ (a
$4\pi$ per interaction assuming the theory becomes strongly coupled at
$M_*$).  Thus the relevant part of the Lagrangian is given by
\begin{eqnarray}
  \label{eq:17}
   {\cal L}={(4\pi)^2\over M_*^4} \left(QD_M dd\gamma^M\bar
  L + d\,D_M\,dd\gamma^M\bar l \right)+\dots \,,\label{L} 
\end{eqnarray}
where $M_*$ is the effective cutoff introduced in section
\ref{sec:framework}.  Again, using NDA we estimate the rate for B$-$L
violation outside the bubble, $\Gamma_{\rm B-L}$ to be roughly 
\begin{equation}
  \label{eq:28}
  \Gamma_{\rm B-L}\sim \frac{(4\pi)^4}{8\pi} \,\left({T \over M_*}\right)^8 \,T \,.
\end{equation}
In order for the interaction to be in
equilibrium we require it to be faster than the expansion rate at the
critical temperature,
\begin{equation}
  \label{eq:35}
   \Gamma_{\rm B-L}\sim \frac{(4\pi)^3}{2}\, T_{\rm LPT}\, \left({T_{\rm LPT}\over
    M_*}\right)^8\geq {T^2_{\rm LPT} \over M_{\rm Pl}}\,,
\end{equation}
where $M_{\rm Pl}$ is the reduced Planck mass.  Using, $T_{\rm LPT}
\sim 1/\pi R$, this
yields the following condition
\begin{equation}
  \label{eq:40}
   {2\over (4\pi)^3}\,\, {\left({\pi R
      M_*}\right)^7}\,{M_*\over M_{\rm Pl}}\leq 1\,.
\end{equation}
The above condition is easily satisfied over a wide range of $M_*$ and
$R$.  For example for $R M_*=\lambda_C^{-1}$ as long as $M_*\leq
10^{13}\,$GeV the above process is in thermal equilibrium outside the
bubble.  Interestingly, this yields an upper bound for the theory's
cutoff and inverse radius in our model.  The exponentially suppressed
overlaps between the lepton and quark fields (and between themselves)
will ensure that the above processes decouple inside the broken phase.

Finally, we need to estimate the suppression of the baryon number
production which is induced by these processes. This suppression
occurs due to the fact that the typical time scale related to the
bubble wall is expected to be of the order of $t_{\rm w} \sim 4/T_{\rm
  LPT} \,v_{\rm w}$~\cite{Nelson:1991ab,HT}.  Thus if $\Gamma_{B-L}
t_{\rm w}\ll 1$ the B$-$L processes are not efficient enough in
producing baryon number near the wall, before the corresponding
region is overtaken by the bubble.  We, therefore, expect that the
baryon asymmetry to be suppressed by a factor $\epsilon_{\rm B-L}$
given by
\begin{equation}
  \label{eq:41}
  \epsilon_{\rm B-L}\sim \Delta_{\rm B-L}\,\Gamma_{\rm B-L}\, t_{\rm w}\sim
 4 \frac{(4\pi)^3}{v_{\rm w}}\,\left({1\over \pi RM_*}\right)^8
 \sim \frac{1}{v_{\rm w}}\, \left(RM_*\right)^{-8}\,,
\end{equation}
where $\Delta_{\rm B-L} = 2$ is the B$-$L charge difference induced by a
single B$-$L violating process (\ref{eq:12}).

\section{The CP Asymmetry}
\label{sec:cp-asymmetry}

The condition for CP conservation in the SM can be formalized by the
existence of a matrix $K_L$ such that
\begin{eqnarray}
  \label{eq:33}
  K_L^\dagger Y_uY_u^\dagger K_L &=& (Y_uY_u^\dagger)^*,
  \nonumber \\
  K_L^\dagger Y_dY_d^\dagger K_L &=& (Y_dY_d^\dagger)^*.
\end{eqnarray}
In the five dimensional theory at hand, the couplings with the
localizer (see eq. \eqref{eq:37}) add
the condition (considering for a moment only the
quark-doublets)\footnote{In principle this theory contains ten CPV
  phases~\cite{APS}.
Here we only consider for simplicity the doublet sector which contains
three such new phase. As shown below our mechanism is successful even
if all the new phases are switched off which can be achieved
naturally as we argue in section~\ref{sec:conclusions}. }
\begin{eqnarray}
  \label{eq:34}
  K_L^\dagger f_Q K_L = f_Q^*\,.
\end{eqnarray}
As we now show, CP violation in our cosmological context is
found (perturbatively) to be proportional to ${\rm Tr}[f_Q, Y_uY_u^\dagger +
Y_dY_d^\dagger]^3$. Generically we expect it to be of order one.

We calculate the CP
asymmetry in two different limits as follows:
\begin{itemize}
\item Perturbative three-flavor limit, $u/\sqrt{M_*} \ll
1/2\pi R$ - where $u$ is the localizer's (roughly constant) VEV. This would
allow for an analytical computation of the
CP asymmetry within a full three generation framework.
\item Non-perturbative single flavor limit, $u/\sqrt{M_*} \gsim 1/2\pi
  R$ - an
 analytic solution is found for the reflection amplitude within
 a single flavor framework.
\end{itemize}
The results of our analysis in both limits shows that the CP asymmetry
is expected to be sizable.  Generically, the parameters of our model
lie somewhere between the above limits.  The exact result is then
understood to have order one corrections, but is not expected to be
dramatically changed.

Before moving to the actual analysis let us explain how each of the
above limits captures different essential ingredients of the dynamics
of the interactions with the bubble wall. Such an analysis can, in
fact, be useful also in other baryogenesis models beyond our framework
(see {\it e.g}~\cite{Berkooz:2004kx,NP1,HalMurPer}).  In order to
explain the low energy values of the Yukawa couplings, it is required
from the quark odd masses to be few times larger than
1/R~\cite{KT}. In our framework this implies that the localizer's VEV
should be roughly $u \sim 10/\pi R \sqrt{M_*}$ \cite{Grossman:2002pb}.
On the other hand, as discussed before, we assume that the critical
temperature is of order $T_{\rm LPT} \sim 1/\pi R$.  For the above
value of $u$, given that the quarks typical energy is of the order of
the critical temperature, the reflection of the quarks from the
localizer wall cannot be treated perturbatively. This implies that the
non-perturbative treatment can only serve as a rough estimation for
the resultant CP asymmetry.  On the other hand the non-perturbative
limit can be solved only within a single flavor framework and
therefore describes a CP conserving model. It can only be used to
estimate the reflection/transmission probability and not the
reflection asymmetries.

\subsection{The Effective Dirac Equation}
\label{Effective Dirac Equation}

Considering a $2+1$ dimensional problem, during the phase transition
$\phi$ obtains an $x^5$ dependent VEV and the
bubble of true vacuum is expanding in the $x^3$ direction.  Hence the
relevant part of the Lagrangian takes the form
\begin{equation}
  \label{eq:5}
  {\cal L}_5 = i\bar\Psi_i \gamma^M\partial_M\Psi_i -  f_{ij}(x^3,x^5)\bar\Psi_i\Psi_j\,,
\end{equation}
where within the thin wall approximation, 
\begin{equation}
  \label{eq:72}
  \f(x^3, x^5) = \left\{
    \begin{array}{lll}
      0 && x^3\leq 0 \cr
      \left(u/M_*^{1/2}\right)\f && x^3>0
    \end{array}\right..
\end{equation}
To demonstrate how the
asymmetry is being produced, we consider only the quark doublets, namely $f=f^Q$.  The
computation of the CP asymmetry for the other quarks follow in the
same manner and the total asymmetry is obtained through the
thermalization of the doublets and the singlets due to Yukawa
interactions.  Written in matrix form, it is apparent that the spin-up
and spin-down parts of the Dirac fermion decouple,
\begin{eqnarray}
  \label{eq:75}
   {\cal L}_5 =&
    \mypmatrix{
      {\Psi^1_+}^\dagger & {\Psi^3_-}^\dagger}
    \mymatrix{
      E+i\partial_3 & -f(x^3,x^5)-\partial_5 \cr
      -f(x^3,x^5)+\partial_5 & E-i\partial_3}
    \mymatrix{
      \Psi^1_+ \cr
      \Psi^3_-}
    \\
    &+  \mypmatrix{
      {\Psi^2_+}^\dagger & {\Psi^4_-}^\dagger\cr}
    \mymatrix{
      E-i\partial_3 & -f(x^3,x^5)-\partial_5 \cr
      -f(x^3,x^5)+\partial_5 & E+i\partial_3}
    \mymatrix{
      \Psi^2_+ \cr
      \Psi^4_-}
\end{eqnarray}
where the $\pm$ subscripts stands for the 4D chirality and we have
suppressed flavor indices.

In order to compute the CP asymmetry, one must take into account
fermionic interactions with the plasma.  To do so, one introduces the
notion of quasiparticles~\cite{Weldon:1982aq,Farrar:1993sp,Farrar:1993hn,Huet:1994jb}, which
are fermionic excitations in the plasma.  As we review in Appendix
\ref{sec:effect-dirac-equat}, the effective Dirac equation is found to
be
\begin{eqnarray}
 \label{eq:6}
  \mymatrix{
    \p+i\partial_3 + i/2l & -{\cal F}(x^5)\theta(x^3)-\partial_5 \cr
    -{\cal F}(x^5)\theta(x^3)+\partial_5 & \p-i\partial_3 + i/2l}\chi(x^3, x^5)
  &=& 0,
  \\
  \label{eq:7}
  \mymatrix{
    \p-i\partial_3+i/2l & -{\cal F}(x^5)\theta(x^3)-\partial_5 \cr
    -{\cal F}(x^5)\theta(x^3)+\partial_5 & \p+i\partial_3+i/2l}\tilde\chi(x^3,x^5) &=&0 ,
\end{eqnarray}
where  $l$ is the coherence length of the quasiparticles given by $l \simeq
1/g_s^2T$ \cite{Braaten:1992gd,Huet:1994jb}. Thus the quasiparticles
are damped as they propagate in space.  We also define,
\begin{equation}
  \label{eq:9}
  \chi \equiv  \mymatrix{
    \Psi^1_+ \cr
    \Psi^3_-}; \;\;\;
  \tilde\chi \equiv \mymatrix{
    \Psi^2_+ \cr
    \Psi^4_-}  
\end{equation}
and
\begin{eqnarray}
  \label{eq:10}
  {\cal F}(x^5) &\equiv& (3u(x^5)/2M_*^{1/2})f,
  \\
  \label{eq:27}
  \p  &\equiv&  3(E-\Omega).
\end{eqnarray}
Here $\Omega$ denotes the thermal masses of the quasiparticles which
are given, to leading order, by \cite{Weldon:1982aq,Farrar:1993sp,Huet:1994jb},
\begin{eqnarray}
  \label{eq:3}
  \Omega_Q^2 &=& \frac{2\pi\alpha_s T^2}{3} +
  \frac{\pi\alpha_WT^2}{2}\left(\frac{3}{4} +
    \frac{\sin^2\theta_W}{36}\right) + \frac{T^2}{16}\left(Y_uY_u^\dagger +Y_dY_d^\dagger\right),
  \\
  \Omega_u^2 &=& \frac{2\pi\alpha_s T^2}{3} +
  \frac{2\pi\alpha_W\sin^2\theta_WT^2}{9} + \frac{T^2}{4}Y_u^\dagger Y_u,
  \\
  \Omega_d^2 &=& \frac{2\pi\alpha_s T^2}{3} +
  \frac{2\pi\alpha_W\sin^2\theta_WT^2}{9}+  \frac{T^2}{4}Y_d^\dagger Y_d.
\end{eqnarray}
Note that here $Y_u$ and $Y_d$ are order one couplings as
opposed to the four dimensional case.  Here we have neglected
contributions related to the localizer couplings which, although introduce further
flavor dependence, do not play a role in our scenario.

We next decompose
the fields into KK modes so that the above equations acquire the form
similar to the 4D case.  This is done in detail in appendix
\ref{sec:effect-dirac-equat}.  We find,
\begin{eqnarray}
  \label{eq:30}
  \mymatrix{
    \p+i\partial_3+i/2l & -\theta(x^3)\F^{mn}-p_5^{(n)} \\
    -\theta(x^3)\F^{Tmn}-p_5^{(n)} & \p-i\partial_3+i/2l}\chi^{(n)}(x^3)
  &=& 0,
  \\
  \label{eq:30b}
  \mymatrix{
    \p-i\partial_3+i/2l & -\theta(x^3)\F^{mn}-p_5^{(n)} \\
    -\theta(x^3)\F^{Tmn}-p_5^{(n)} & \p+i\partial_3+i/2l}\tilde\chi^{(n)}(x^3) &=& 0.
\end{eqnarray}
Here, $p_5^{(n)} = n/R$, we have suppressed flavor indices, summation
over KK indices is implied and
$\F^{nm}$ is given by,
\begin{eqnarray}
  \label{eq:47}
  \F_{ij}^{mn} = \left\{\begin{array}{lllll}
        \frac{6u}{\pi M_*^{1/2}}\frac{n}{n^2-m^2}\f&&&& m-n\in
        2Z+1 \\
        0 &&&& \mathrm{Otherwise}
        \end{array}\right..
\end{eqnarray}
We remind the reader that for quark doublets (singlets),
$\Psi^{(0)}_-$ ($\Psi^{(0)}_+$) is projected out.

\subsection{A Perturbative Approach}
\label{Dirac:Perturbative}
We now turn to solve eq. \eqref{eq:30} perturbatively, assuming $\F^{mn}$
is small.  This allows us to find a quantitative estimate for the size
of the induced asymmetry.  Since the critical temperature is of the
order of inverse the 5D radii, it is enough, when calculating the
transmission and reflection coefficients, to consider only the first
few KK modes.  All higher modes are unexcited due to a strong
exponential Boltzmann suppression.  In practice we shall only consider
the zero and the first KK states (note that for each fermion there is a single
zero mode and two first KK modes with different helicities).

As is clear from the Green's functions described below, only $\chi$
has an incoming zero-mode.  Hence considering $\chi$ and adding a
source for the incoming zero-mode, eq. \eqref{eq:30} becomes,
\begin{eqnarray}
  \label{eq:2}
 &&   \mymatrix{
    \p_{ij}+i\partial_3+i/2l & - p_5^{(n)} \\
     -p_5^{(n)} & \p_{ij}-i\partial_3+i/2l}
  \mymatrix{
    \Psi_{j+}^{1(n)}(x^3) \\
    \Psi_{j-}^{3(n)}(x^3)}=\nonumber\\
  &&
  \mymatrix{
    -i\delta(x^3)\delta_{ij}\delta^{0m}\delta^{n0} & \theta(x^3){\cal F}_{ij}^{nm} \\
    -\theta(x^3){\cal F}_{ij}^{Tnm} & 0
  }
  \mymatrix{
    \Psi_{j+}^{1(m)}(x^3) \\
    \Psi_{j-}^{3(m)}(x^3)
  }.
\end{eqnarray}
This equation describes free fermions with mass $p_5^{(n)}$
propagating in the unbroken phase, taken to be at $x^3 < 0$.  As
fermions hit the wall, they interact with the localizer which changes
their chirality and partially reflect them back.  We treat $\F^{mn}$
perturbatively, describing quasiparticles which reflect back and forth
inside the broken phase.  Due to the non-diagonal form of $\F^{mn}$ in
KK space, an incoming zero-mode is reflected back into an odd KK-mode
(see eq. \eqref{eq:47}).
For the reasons mentioned above we consider only an additional first
excited KK mode, and so we seek an expression for the reflection
coefficients, describing the scattering of zero-mode into a first KK
mode.  In order to be sensitive to CP violation, we must take into account
multiple scattering so as to obtain interference.

The free Green's function for this problem is
constructed in Appendix \ref{Green's Function}.  We find,
\begin{eqnarray}
  \label{eq:11}
   \hat G^{(n)}(x^3-x^{3\prime}) = &&-\mymatrix{
    \frac{\p+p_3^{(n)}}{2p_3^{(n)}} & \frac{p_5^{(n)}}{2p_3^{(n)}} \\
    \frac{p_5^{(n)}}{2p_3^{(n)}}& \frac{\p-p_3^{(n)}}{2p_3^{(n)}}
  }\theta(x^3-x^{3\prime})e^{ip_3^{(n)}x^3}\nonumber\\
  && - \mymatrix{
    \frac{\p-p_3^{(n)}}{2p_3^{(n)}} & \frac{p_5^{(n)}}{2p_3^{(n)}} \\
    \frac{p_5^{(n)}}{2p_3^{(n)}}& \frac{\p+p_3^{(n)}}{2p_3^{(n)}}
  }\theta(x^{3\prime}-x^3)e^{-ip_3^{(n)}x^3},
\end{eqnarray}
where $p_3^{(n)} = \sqrt{\p^2-(p_5^{(n)})^2} + i/2l$ is flavor dependent
through $\p$ and is KK-dependent though $p_5^{(n)}$. Using the Green's function one
easily finds the reflection coefficient.  We do this in Appendix
\ref{A Perturbative Computation of R}.  For now we give the result
and note that as opposed to the usual 4D case where in order to see CP
violation one must expand to fifth order in perturbation theory, here
CP violation is apparent already at third order.  Indeed,
\begin{eqnarray}
  \label{eq:14}
    R_{01} &=& i\mymatrix{
    G^{(1)}_{12}\F G_{11}^{(0)} \left[1 - \F G_{22}^{(1)}\F
      G_{11}^{(0)}\right]
    \\
    G^{(1)}_{22}\F G_{11}^{(0)} \left[1 - \F G_{22}^{(1)}\F
      G_{11}^{(0)}\right]
  }.
\end{eqnarray}
where here $\F = \F^{01}g_s^{-2}2\pi R = (12uR/ g_s^2 M_*^{1/2})f$ and
the sub-indices refer to chirality space.

Plugging the above reflection coefficient in the CP-odd expression
$\Delta_{\rm CP}^{nm}$ given in \eqref{eq:43} and expanding the
Green's function in the flavor dependent part, one finds (see Appendix \ref{A Perturbative
  Computation of R}),
\begin{eqnarray}
  \label{eq:46}
   \Delta_{\rm CP}^{01} &\propto& l^3\frac{1}{3}{\rm
   Tr}\left(\F\right){\rm tr}\left([\F,\delta\p]^3\right) = l^3{\rm
   Tr}\left(\F\right)\det\left([\F,\delta\p]\right) \simeq 10^{-2}
\end{eqnarray}
with $\delta\p_Q = (T/32)(Y_uY_u^\dagger + Y_dY_d^\dagger)$ and
$\delta\p_{u,d} = (T/8)Y_{u,d}Y_{u,d}^\dagger$.  This result, as
advocated, has only mild numerical suppressions arising from the loop
corrections to the thermal masses.

\subsection{A Non-perturbative Approach}
\label{Nonpert}
The reflection coefficients may be calculated in a very different way.
Instead of the perturbative approach which is sensitive to the CPV in
the theory we can solve a simpler reflection
problem semi-analytically.  This is important since in order to solve the flavor
puzzle we require $f$ of order a few which is beyond the perturbative
regime. The problem with the following approach is that the
calculation can only be done in a single flavor framework so CP
violation is absent. Our attempt here is therefore only to have a
quantitative estimate for the magnitude of the reflection
coefficients.  We then assume that when promoting the problem to
accommodate three generations, the reflection coefficients become
complex while their magnitudes only change by order one coefficients.
The explicit calculation is done in Appendix \ref{sec:non-pert-calc-1}
and here we only summarize our results.

\begin{figure}[!t]
\begin{center}
\includegraphics[height=6cm, width=9cm]{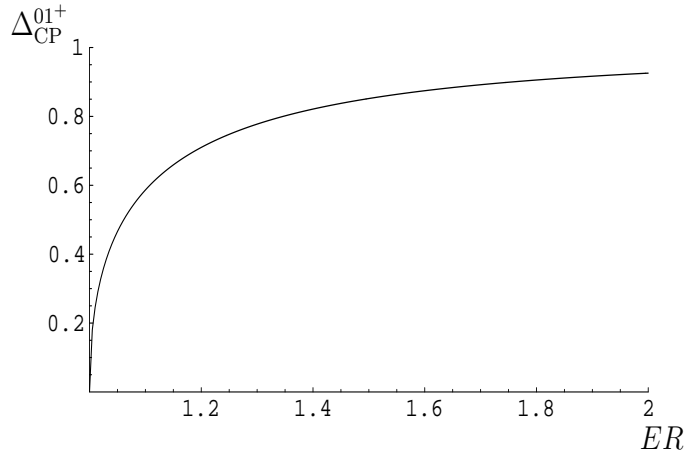}
\caption{Estimation of $\Delta(E)^{01^+}$, the probability of a zero
  mode to be reflected into a positive helicity state, in a single flavor model for $1<ER<2$.
}\label{figRE}
\end{center}
\end{figure}

For the calculation we adopt the following strategy.  We apply the
thin wall approximation and solve the Dirac equation exactly inside
and outside the bubble for the whole KK tower.  Since the dominant
contributions come from the lowest KK modes, we consider only the
zeroth and first KK modes as on shell modes ({\it i.e.} only these
modes will contribute to the reflection and transmission asymmetry).
We then match the wave functions of the incoming zero mode and the
first KK mode outside the bubble to the outgoing ones inside the
bubble.  This allows us to extract the reflection coefficients.  While
this procedure is clear and analytic, the final stage of the actual
matching involves the matching of a single state outside the bubble to
an infinite tower of (virtual) states inside the bubble which can only
be done numerically.  We verified our numerical results in several
limiting cases as described in the Appendix.  Our main result is that
a sizable reflection is found over a wide range of
energies~\cite{NP1}.  This is shown in fig. \ref{figRE} which shows
the reflection coefficient for an incoming zero mode.  It is apparent
from the figure, that the reflection coefficients vanish as the energy
drops below $1/R$.  This is expected since at such energies only the
zero-modes are excited, and thus cannot be reflected by the wall. In
our analysis we only consider the first KK states so that it applies
only for $w<2/R\,.$

As a final remark, we list the correct unitarity relations between
the reflection and transmission coefficients, which are
calculated in Appendix \ref{apUnit}. 
\beq
r_k |R_{01^+}|^2+|T_{00}|^2=1\,,\ \ \
|R_{1^-1^+}|^2+{1\over r_k}|T_{1^-0}|^2=1\,,\ \ \
 {1\over r_k}|R_{1^+0}|^2+|R_{1^+1^-}|^2=1\,,
\label{Unit}
\eeq
where $r_k=\sqrt{1-{1\over E^2 R^2}}$ and we have
explicitly written the helicities of the incoming and outgoing states.  As was remarked
in section \ref{sec:cosmological-setup}, these exact relation would
imply that the
final expression for $n_B/s$ is slightly altered when the realistic
values are considered for the model's fundamental
parameters. Nevertheless our-order-of-magnitude estimate
for the asymmetry is insensitive for such order one effects.

\section{Discussion \& Conclusions}\label{sec:conclusions}

We begin this part by presenting our final result.  In section
\ref{sec:bary-numb-viol} we showed that, $\epsilon_{\text{B-L}}$, the
suppression from the inefficiency of the B$-$L production near the
bubble wall is given by $\epsilon_{\text{B-L}}\sim 1/[v_{\rm w}
(RM_*)^8]$. In section \ref{sec:cp-asymmetry} we find that,
$\Delta_{\rm CP}^{01}$, the CP asymmetry in the reflection rates is
mildly suppressed and is given by $\Delta_{\rm CP}^{01} \sim 10^{-2}$.
Boltzmann suppression and other numerical factors provide another
suppression of order ${\cal O}(10^{-1})$. We therefore estimate 
the ratio between the baryon asymmetry and entropy to be given by
\begin{equation}
  \label{eq:44}
  \frac{n_B}{s}\sim {10^{-1}\over g_*}\; \epsilon_{\text{B-L}}\; \Delta_{\rm
 CP}^{01}
\sim 10^{-10}\times \left(\lambda_{C}\times  RM_*\right)^{-8}.
\end{equation}
where $g_*\sim10^2$ and $\lambda_C\sim 0.22$ is the Cabibbo mixing angle
and for comparison with Froggatt-Nielsen type models it is convenient to
consider $1/ RM_*$ in units of $\lambda_C\,.$ This is since in the Froggatt-Nielsen
framework $\lambda_C$ sets the ratio between the Froggatt-Nielsen scalar
VEV and the effective cut-off scale.
Thus our estimation for the baryon asymmetry is consistent with the
observed value.

The flavor puzzle of the standard model (SM) is related to the fact
that most of the flavor parameters are small and hierarchical which
hints towards physics beyond the SM.  Several kinds of solutions have
been suggested to solve the above puzzle.  They all require additional
fields with new dynamics.  Our main message in this work can be stated
as follows: In a cosmological context the additional dynamics, which
exists in flavor models, naturally incorporates order one Yukawa
couplings in the early universe.  This could result with a baryon
asymmetry produced with the KM phase as the only source of CPV.  Thus it is
plausible that the flavor puzzle and the SM failure to create the
baryon asymmetry are linked and have a common resolution which does
not require the presence of new CP violating sources.  This is not to
say that no new CP violating sources exist beyond the SM.  We merely point out
that the KM phase may play an important role in the production of the
baryon asymmetry.

In~\cite{Berkooz:2004kx} the above concept was introduced and realized
in Froggatt-Nielsen type models in four dimensions.  In this paper we
illustrate how a similar behaviour is obtained within a different
framework which solves the flavor puzzle, that is the split fermion
framework.  Assuming a first order phase transition in which a field
(the localizer) responsible for the localization obtains a VEV, a net
baryon number is produced.  CPV occurs through
interactions of fermions with the bubble wall.  Furthermore
non-renormalizable B$-$L violating interactions, which are allowed by
the SM symmetries, induce baryon production outside the bubble wall. These
interactions decouple in the broken phase due to fermions localization
and thus play a similar role to the sphaleron processes in the SM case.

We argued above that no new sources of CP violation are required for
our mechanism to work.  In our model CP violation was found to be
proportional to (at the perturbative level) ${\cal J}_f={\rm
  Tr}(f){\rm det}[f, Y^\dagger Y]$ where $f$ and $Y$ denote the Yukawa
couplings of the fermions to the localizer and the Higgs respectively.
Both $f$ and $Y$ are matrices in flavor space and are expected to be
anarchical (we omit the flavor and representation indices for
simplicity). Let us demonstrate that indeed one can construct a
natural model in which the KM phase is the only CP violating source.
Consider a model in which the two Yukawa matrices are promoted to be
fields which transform as bi-fundamentals of the corresponding SM
$U(3)_Q\times U(3)_u\times U(3)_d$ flavor group.  In the case in which only these two
bi-fundamental fields break the flavor group the non-universal part of
$f$ will be given, to leading order, by a bilinear combination of the
Yukawas. Thus it is clear that any CP violating quantity in the above
theory (in particular ${\cal J}_f$) must be proportional to the
commutator of the two bi-fundamental fields. Thus the only CP
violating source is the KM phase.  This should hold even in the presence of quantum
correction and therefore technically natural.

In~\cite{NP1} it was demonstrated that realistic split fermion
models may induce lepton production from different CP violating
sources~\cite{HPSS}, namely through twisting of the fermion wave function in flavor
space~\cite{GHPSS}.
It is interesting that the effect described above is complimentary to
the one induced via twisting and in general both are expected to
be present.

Our mechanism requires that the critical temperature should be of the
order of the inverse compactification scale, $1/R$.  Phenomenological
constraints typically yield $1/R \gsim 100\,$TeV.  This implies that
our framework allows for a rather low reheating temperature, after
inflation, solving the various moduli/gravitino problems.  Finally, we
note that our model which is based on Kaplan-Tait type
models~\cite{KT}, with a constant localizer VEV, is compatible with
five dimensional supersymmetric split fermion models~(see {\it
  e.g.}~\cite{SUSY}). Hence it can easily be supersymmetrized.

\section{ Acknowledgments }
The authors thank Yaron Antebi, Guy Engelhard, Guy Raz and Ze'ev
Surujon for useful discussions and especially Yuval Grossman, Antonio
Masiero and Yossi Nir for useful discussions and comments on the
manuscript.  GP thanks the Aspen Center for Physics and the Institute
of Theoretical Physics, Technion for hospitality while this work was
in progress.  GP is supported by DOE under contract DE-AC03-76SF00098.



{\bf {\Large~~~~~~~~~~~~~~~~~~~~~~~ APPENDIX}}

\appendix

\section{The Effective Dirac Equation and KK Decomposition}
\label{sec:effect-dirac-equat}

As explained in \cite{Huet:1994jb}, the dispersion relation for 
quasiparticles may be found by looking for the poles of the full
propagator, including the self-eneries.  Taking thermal self-energy of
the form,
\begin{equation}
  \label{eq:20}
  \Sigma(E, \vec p) = \gamma^0a(E, p)-b(E,p)\vec\gamma\cdot\vec p,
\end{equation}
one is then looking for a solution to
\begin{equation}
  \label{eq:21}
  \det[S_0^{-1}-\Sigma(E, \vec p)] = 0,
\end{equation}
with $S_0^{-1} = \gamma^0E-\vec\gamma\cdot\vec p$.  Considering the
$2+1$ problem at hand, the solution to the above is
\begin{equation}
  \label{eq:23}
  E = a(E,p) \pm \left[1-b(E,p)\right]\sqrt{p_3^2+p_5^2}.
\end{equation}
In the four dimensional case, the self-energy can be linearized as
\begin{equation}
  \label{eq:24}
  \Sigma(E,\vec p)\simeq \gamma^0(2\Omega-i/3l-E) - \vec\gamma\cdot\vec p/3,
\end{equation}
where as mentioned in section \ref{Effective Dirac Equation}, $\Omega$
denotes the thermal mass of the quasiparticle and $l$ is the coherence
length of the quasiparticles given by $l = 1.1/g_s^2T$
\cite{Braaten:1992gd,Huet:1994jb}.  This is of course also true in the
broken phase where the theory is effectively 4D.  In the symmetric
phase, the theory is effectively 5D and thus we expect corrections to
the above.  Nevertheless, at the temperature of phase transition, the 5D
linearized self-energies are expected to be affected only by a few
higher KK states and are thus expected, at zeroth order, to admit the
same behaviour.  With the above, one obtains the dispersion relation,
\begin{equation}
  \label{eq:25}
  E = \Omega -\frac{i}{6l}\pm \frac{1}{3}\sqrt{p_3^2+p_5^2}.
\end{equation}
The effective free Dirac equation is therefore obtained by taking
$\gamma^\mu p_\mu \rightarrow \gamma^\mu p_\mu-\Sigma$ which has the
effect of taking $E \rightarrow 2(E-\Omega+i/6l)$ and $p \rightarrow
2p/3$.  Thus, plugging these in eq. \eqref{eq:75} and massaging a
little one finds the effective Dirac equation given in
\eqref{eq:6},\eqref{eq:7}.

Next we KK decompose the Dirac fermions.  The KK reduction of the wave
function is,
\begin{eqnarray}
  \label{eq:8}
  \Psi_{i\pm}(x^3, x^5) = \sum_{n=0}^\infty u_{in}^{\pm}(x^5)\Psi_{i\pm}^{(n)}(x^3),
\end{eqnarray}
where the $\pm$ are the two chiralities of the Dirac fermion, and
$u_{in}^\pm = a_{in}^\pm e^{ip_5x^5} + b_{in}^\pm e^{-ip_5x^5}$.  The
boundary conditions for $u_{in}^\pm$ are given by (again, considering, for
example, the quark doublets),
\begin{eqnarray}
  \label{eq:16}
  u_{in}^\pm(x^5=0^+,\pi R^-) &=& \pm u_{in}^\pm(x^5=0^-,\pi R^+)\,,
  \\
  \partial_5 u_{in}^+|_{x^5=0^+,\pi R^-}&=& \f(x^5=0^+,\pi
  R^-)u_{jn}^+(x^5=0^+,\pi R^-)\,,
  \\
  \partial_5 u_{in}^-|_{x^5=0^+,\pi R^-}&=&\partial_5 u_{in}^-|_{x^5=0^-,\pi R^+}.
\end{eqnarray}
Hence in the unbroken phase,
\begin{eqnarray}
  \label{eq:4}
  u_{in}^+ = \left(\frac{2}{\pi R}\right)^{1/2}\cos\left(\frac{nx^5}{R}\right), \hspace{1cm} u_{in}^- =
  \left(\frac{2}{\pi R}\right)^{1/2}\sin\left(\frac{nx^5}{R}\right).
\end{eqnarray}

It is important to note that under this decomposition, the thermal
masses are KK diagonal.  To see this simply note that we assume the
Higgs field is $x^5$ independent.  Thus the Yukawa interactions decompose into
\begin{eqnarray}
  \label{eq:19}
  Y_{ij}H \bar Q d \longrightarrow
  Y_{ij}H\sum_n\left(\bar Q_+^{(n)}d_-^{(n)}+ \bar Q_-^{(n)}d_+^{(n)}\right).
\end{eqnarray}
Here we replaced (temporarily) $\Psi$ by $Q$ and $d$ to stress that
the interaction is between two distinct fields.  Note that $Q$ and $d$
have orbifold boundary conditions with opposite sign.  The above in
particular means that $\Omega_\pm^{(n)}$ are independent of $n$.
Moreover, $\Omega_+ = \Omega_-$.

On the other hand, the interactions with the localizer are not KK
diagonal.  We have,
\begin{eqnarray}
  \label{eq:22}
  &&f_{ij}(x^3,x^5)\bar\Psi_i\Psi_j \longrightarrow \theta(x^3)\sum_{nm} \frac{2}{3}\left({\cal
  F}_{ij}^{mn}\bar\Psi_{i+}^{(m)}\Psi_{j-}^{(n)} + {\cal
  F}_{ij}^{Tmn}\bar\Psi_{i-}^{(m)}\Psi_{j+}^{(n)}\right),
\end{eqnarray}
with $\F$ given in eqs. \eqref{eq:10},~\eqref{eq:47}.  In the above, 
the transpose is with respect to KK number.  Note further, that due to the
complex conjugate term in the Lagrangian, ${\cal F}$ can be taken
to be Hermitian with respect to flavor.

\section{Computation of the Green's Function}
\label{Green's Function}

To construct the (free) Green's functions we solve:
\begin{eqnarray}
  \label{eq:66}
  \mymatrix{
    ({\cal P_+})_{ij}+i\partial_3 & -p_5^{(n)} \\
    -p_5^{(n)} & ({\cal
    P_-})_{ij}-i\partial_3}\hat G_{jk}^{(n)}(x^3-x^{3\prime})
  &=& i\delta(x^3-x^{3\prime})\delta_{ik},
  \\
  \label{eq:67}
   \mymatrix{
    ({\cal P_+})_{ij}-i\partial_3 & -p_5^{(n)} \cr
    -p_5^{(n)} & ({\cal
      P_-})_{ij}+i\partial_3}\hat{\tilde G}_{jk}^{(n)}(x^3-x^{3\prime})
  &=& i\delta(x^3-x^{3\prime})\delta_{ik},
\end{eqnarray}
where $\hat G_{jk}^{(n)}(x^3-x^{3\prime})$ is a $2\times 2$ matrix.  The
solution is found by a double Fourier transform.  Defining,
\begin{eqnarray}
  \label{eq:68}
  \hat G_{jk}^{(n)}(x^3-x^{3\prime}) = \int dk e^{ik (x^3- x^{3\prime})}
  \hat G_{jk}^{(n)}(k)
  \\
  \hat{\tilde G}_{jk}^{(n)}(x^3-x^{3\prime}) = \int dk e^{ik (x^3- x^{3\prime})}
  \hat{\tilde G}_{jk}^{(n)}(k)
\end{eqnarray}
we obtain (using $\p_+ =\p_- \equiv \p$),
\begin{eqnarray}
 \label{eq:69}
  \hat G_{jk}^{(n)}(k) &=& 2\pi i\mymatrix{
    \p-k & -p_5^{(n)} \\
    -p_5^{(n)} & \p+k}^{-1}_{jk} = -2\pi i\mymatrix{
    \frac{k+\p}{k^2-(p_3^{(n)})^2} &
    \frac{p_5^{(n)}}{k^2-(p_3^{(n)})^2} \\
     \frac{p_5^{(n)}}{k^2-(p_3^{(n)})^2} &
     \frac{-k+\p}{k^2-(p_3^{(n)})^2}
   }_{jk} 
  \\
  \label{eq:70}
  \tilde{\hat G}_{jk}^{(n)}(k) &=& 2\pi i\mymatrix{
    \p+k & -p_5^{(n)} \\
    -p_5^{(n)} & \p-k}^{-1}_{jk}= -2\pi i\mymatrix{
    \frac{-k+\p}{k^2-(p_3^{(n)})^2} &
    \frac{p_5^{(n)}}{k^2-(p_3^{(n)})^2} \\
     \frac{p_5^{(n)}}{k^2-(p_3^{(n)})^2} &
     \frac{k+\p}{k^2-(p_3^{(n)})^2}
   }_{jk},
 \end{eqnarray}
 where $(p_3^{(n)})^2 = \p^2-(p_5^{(n)})^2$.
This is precisely the Green's function
of a massive fermion, $(\gamma_\mu p^\mu +
m)/(p^2-m^2)$ with $m = p_5^{(n)}$.  The inverse Fourier transform depends on the contour chosen for the
integral which is equivalent to choosing a boundary
condition.  There are four different contours going above or below
each of the two poles $k = \pm p_3^{(n)}$.  For reasons to be
understood below, we choose the contour which goes {\bf above}
$k=-p_3^{(n)}$ and {\bf below} $k=+p_3^{(n)}$.  After a Cauchy integral one
obtains,
\begin{eqnarray}
  \label{eq:73}
  \hat G_{jk}^{(n)}(x^3-x^{3\prime}) &=& -\mymatrix{
    \frac{\p+p_3^{(n)}}{2p_3^{(n)}} & \frac{p_5^{(n)}}{2p_3^{(n)}} \\
    \frac{p_5^{(n)}}{2p_3^{(n)}}& \frac{\p-p_3^{(n)}}{2p_3^{(n)}}
  }_{jk}\theta(x^3-x^{3\prime})e^{ip_3^{(n)}x^3}\nonumber\\
  && - \mymatrix{
    \frac{\p-p_3^{(n)}}{2p_3^{(n)}} & \frac{p_5^{(n)}}{2p_3^{(n)}} \\
    \frac{p_5^{(n)}}{2p_3^{(n)}}& \frac{\p+p_3^{(n)}}{2p_3^{(n)}}
  }_{jk}\theta(x^{3\prime}-x^3)e^{-ip_3^{(n)}x^3},
  \\
  \tilde{\hat G}_{jk}^{(n)}(x^3-x^{3\prime}) &=& -\mymatrix{
    \frac{\p-p_3^{(n)}}{2p_3^{(n)}} & \frac{p_5^{(n)}}{2p_3^{(n)}} \\
    \frac{p_5^{(n)}}{2p_3^{(n)}}& \frac{\p+p_3^{(n)}}{2p_3^{(n)}}
  }_{jk}\theta(x^3-x^{3\prime})e^{ip_3^{(n)}x^3} \nonumber\\&&- \mymatrix{
    \frac{\p+p_3^{(n)}}{2p_3^{(n)}} & \frac{p_5^{(n)}}{2p_3^{(n)}} \\
    \frac{p_5^{(n)}}{2p_3^{(n)}}& \frac{\p-p_3^{(n)}}{2p_3^{(n)}}
  }_{jk}\theta(x^{3\prime}-x^3)e^{-ip_3^{(n)}x^3}.
\end{eqnarray}

The above matrices can be diagonalized.  One finds one left-moving
mode and one right-moving mode.  The diagonal Green's functions are
given by
\begin{eqnarray}
  \label{eq:74}
  \hat G_{jk}^{(n)}(x^3-x^{3\prime}) &=& -\mymatrix{
    \frac{\p}{p_3^{(n)}}\theta(x^3-x^{3\prime})e^{ip_3^{(n)}x^3} & 0 \\
    0 & \frac{\p}{p_3^{(n)}}\theta(x^{3\prime}-x^3)e^{-ip_3^{(n)}x^3}
  }_{jk},
  \\
  \tilde{\hat G}_{jk}^{(n)}(x^3-x^{3\prime}) &=& -\mymatrix{
    \frac{\p}{p_3^{(n)}}\theta(x^{3\prime}-x^3)e^{-ip_3^{(n)}x^3} & 0 \\
    0 & \frac{\p}{p_3^{(n)}}\theta(x^3-x^{3\prime})e^{ip_3^{(n)}x^3}
  }_{jk}
\end{eqnarray}
and are obtained through a change of basis with the diagonalizing matrices,
\begin{eqnarray}
  \label{eq:65}
  U^{(n)} = \theta(x^3-x^{3\prime})\mymatrix{
    p_3^{(n)}+\p & \frac{p_3^{(n)}-\p}{p_5^{(n)}} \\
    p_5^{(n)} & 1
  } + \theta(x^{3\prime}-x^3)\mymatrix{
    -(p_3^{(n)}+\p) & \frac{p_3^{(n)}-\p}{p_5^{(n)}} \\
    p_5^{(n)} & -1
  },
  \\
   \tilde U^{(n)} = \theta(x^{3\prime}-x^3)\mymatrix{
    p_3^{(n)}+\p & \frac{p_3^{(n)}-\p}{p_5^{(n)}} \\
    p_5^{(n)} & 1
  } + \theta(x^3-x^{3\prime})\mymatrix{
    -(p_3^{(n)}+\p) & \frac{p_3^{(n)}-\p}{p_5^{(n)}} \\
    p_5^{(n)} & -1
  }.
  \end{eqnarray}
Note that with this choice there is a well defined $p_5\rightarrow 0$
limit, but the matrices are not orthogonal.  One can choose an
orthogonal matrix by normalizing the eigenverctors.  Note also, that there is
only an incoming (outgoing) zero-mode for $\chi$($\tilde\chi$) since there is no $\chi_-^{(0)}$, $\tilde\chi_-^{(0)}$.
Finally, it is clear that the Green's functions fulfill, 
\begin{equation}
  \label{eq:31}
  G_+(-\infty) = G_-(+\infty) = 0.
\end{equation}

\section{A Perturbative Computation of the CP Asymmetry}
\label{A Perturbative Computation of R}
Our goal is a perturbative solution of the reflection coefficients for
the scattering described by eq. \eqref{eq:2}.  As we argued in section
\ref{Dirac:Perturbative}, the lowest order $su(3)$-invariant
expression is of the form ${\rm Tr}(A^3) = 3\det(A)$ for $A\in su(3)$.
Thus the first term to contribute to the asymmetry, $\Delta_{\rm CP}
\sim R_{01}^\dagger R_{01}$, is of form ${\rm Tr}([\F, Y^\dagger
Y]^3)$ and can therefore arise from fourth order in $\F$.  The
reflection coefficients, should therefore be calculated to third order
in $\F$.

Using the Green's functions, one transforms eq. \eqref{eq:2} into an
intergral one,
\begin{eqnarray}
  \label{eq:26}
  \mymatrix{
    \Psi_{i+}^{1(n)}(x^3) \\
    \Psi_{i-}^{3(n)}(x^3)}
  &=& \mymatrix{
    -iG_+^{(0)}(x^3)\Psi_{+i}^{1(0)}(0)\delta^{n0}\\ 0} +
  \\
  && \int dx^{3\prime}\theta(x^{3\prime})
 \hat G^{(n)}(x^3-x^{3\prime})
  \mymatrix{
    0 & {\cal F}^{nm} \\
    -{\cal F}^{Tnm} & 0}
  \mymatrix{
    \Psi_{k-}^{4(m)}(x^{3\prime}) \\
    \Psi_{k+}^{2(m)}(x^{3\prime})},
\end{eqnarray}
where we suppressed the flavor indices.

The solution is found iteratively,
\begin{eqnarray}
  \label{eq:13}
  &&\mymatrix{
    \Psi_{i+}^{1(n)}(x^3) \\
    \Psi_{i-}^{3(n)}(x^3)
  }^0  =
  \mymatrix{
     -iG_{11}^{(0)}(x^3)\Psi_{+i}^{1(0)}(0)\delta^{n0}
  \\ 0},
  \\ \nonumber\\
  &&\begin{split}
    \mymatrix{
      \Psi_{i+}^{1(n)}(x^3) \\
      \Psi_{i-}^{3(n)}(x^3)
    }^1  =  \int& dx^{3\prime}\theta(x^{3\prime})
    \hat G^{(n)}(x^3-x^{3\prime})
    \mymatrix{
      0 & {\cal F}_{jk}^{n0} \\
      -{\cal F}_{jk}^{Tn0} & 0}\mymatrix{
      -iG_{11}^{(0)}(x^{3\prime})\Psi_{+i}^{1(0)}(0)
      \\ 0}
  \end{split},
  \\ \nonumber\\
  &&\begin{split}
    \mymatrix{
      \Psi_{i+}^{1(n)}(x^3) \\
      \Psi_{i-}^{3(n)}(x^3)
    }^2  =  \int
    &dx^{3\prime}dx^{3\prime
      \prime}\theta(x^{3\prime})\theta(x^{3\prime\prime})
    \hat G^{(n)}(x^3-x^{3\prime})
    \mymatrix{
      0 & {\cal F}_{jk}^{nm} \\
      -{\cal F}_{jk}^{Tnm} & 0}
    \hat G^{(m)}(x^{3\prime}-x^{3\prime\prime})
    \\
    &\mymatrix{
      0 & {\cal F}_{jk}^{m0} \\
      -{\cal F}_{jk}^{Tm0} & 0}\mymatrix{
      -iG_{11}^{(0)}(x^{3\prime\prime})\Psi_{+i}^{1(0)}(0)
      \\ 0},
  \end{split}
  \\ \nonumber\\
  &&\begin{split}
    \mymatrix{
      \Psi_{i+}^{1(n)}(x^3) \\
      \Psi_{i-}^{3(n)}(x^3)
    }^3  =  \int
    &dx^{3\prime}dx^{3\prime \prime}dx^{3\prime
      \prime\prime}\theta(x^{3\prime})\theta(x^{3\prime\prime})\theta(x^{3\prime\prime\prime})
    \hat G^{(n)}(x^3-x^{3\prime})
    \mymatrix{
      0 & {\cal F}_{jk}^{nm} \\
      -{\cal F}_{jk}^{Tnm} & 0}
    \\
    &\hat G^{(m)}(x^{3\prime}-x^{3\prime\prime})\mymatrix{
      0 & {\cal F}_{jk}^{ml} \\
      -{\cal F}_{jk}^{Tml} & 0}\hat G^{(l)}(x^{3\prime\prime}-x^{3\prime\prime\prime})\mymatrix{
      0 & {\cal F}_{jk}^{l0} \\
      -{\cal F}_{jk}^{Tl0}& 0}
     \\
     &\mymatrix{
      -iG_{11}^{(0)}(x^{3\prime\prime\prime})\Psi_{+i}^{1(0)}(0)
      \\ 0}.
  \end{split}
\end{eqnarray}
Note that as opposed to the 4D case, here, in general, all orders
contribute to the reflection coefficients due to the non-diagonal form
of the Green's functions.  Nevertheless, the special form of
$\F^{mn}$, eq. \eqref{eq:46}, permits only odd(even) non-vanishing
orders in the expansion for an odd(even) outgoing KK mode.  Even with
this simplification, the general expressions for the reflection
coefficients are very awkward.  To simplify, we consider only the
zeroth and first KK modes.  As explained in section
\ref{Dirac:Perturbative}, this is justified since these two modes have
the lowest Boltzman suppression factors.  Furthermore, numerically,
$\F^{nm}$ is larger for $n,m = 0,1$.  Focusing on these KK
modes, we see that the integrals above are highly damped for $x^3/l
\gg 1$ due to decoherence.  We thus estimate them by replacing $x^3$
with the decoherence length $l = g_s^{-2} 2\pi R$.  Doing so results
with eq. \eqref{eq:14}.

The only flavor dependence in the Green's functions comes from
$Y^\dagger Y$ in $\p$ (by $Y^\dagger Y$ we mean $Y^\dagger_u Y_u
+Y^\dagger_d Y_d$ for the $SU(2)$ doublets and $Y_{u(d)}
Y^\dagger_{u(d)}$ for the singlets). Using $\p \simeq
\p_d + \delta\p$ and $p_3^{(n)} = \sqrt{(\p_d+\delta\p)^2-p_5^{(n)}} +
i/2l$
with $\delta\p \propto Y^\dagger Y$ we expand
\begin{eqnarray}
  \label{eq:29}
  G_{ij}^{(n)} = G_{ij}^{(n)diag}\left(1 + \sum_{k=1}^\infty C_{ij,k}^{(n)}(l\delta\p)^k\right)
\end{eqnarray}
where, $G_{ij}^{(n)diag}$ is the flavor diagonal Green's function
obtained by setting $Y^\dagger Y$ to zero, and for the sake of brevity, we denote the flavor independent
coefficients by $C_{ij,k}^{(n)}$.  Thus for example, $C_{11,k}^{(0)} =
i^k/k!$ etc.

Below we choose the flavor basis such that $\cal F$
is diagonal.  Note that  the diagonalizing matrix does not
contain (by definition) any CP phase, namely, ${\cal F}$ is real.
So to order $\F^4$,
\begin{eqnarray}
  \label{eq:32}
  R^\dagger_{01}R_{01} &=& 
  G_{11}^{(0)\dagger}\F \left(G_{12}^{(1)\dagger}G^{(1)}_{12} +
  G_{22}^{(1)\dagger}G^{(1)}_{22}\right)\F G_{11}^{(0)}\F
  G_{22}^{(1)}\F G_{11}^{(0)}
  \nonumber \\
  &+& G_{11}^{(0)\dagger}\F
  G_{22}^{(1)\dagger}\F G_{11}^{(0)\dagger}\F
  \left(G_{12}^{(1)\dagger}G^{(1)}_{12} +
  G_{22}^{(1)\dagger}G^{(1)}_{22}\right)\F G_{11}^{(0)}
\end{eqnarray}

Each term in eq. \eqref{eq:32} contributes to the CP
asymmetry seperately.  In expanding the Green's functions, the first term to contribute is of
order $\delta\p^3$.  Using the facts that $\delta\p$ is hermitian and
that $\bar R_{01}$ is obtained from $R_{01}$ by replacing $\delta\p$
with $\delta\p^T$  we find, as an example, for the first term in the first line,
\begin{eqnarray}
  \label{eq:18}
  \Delta_{\rm CP} &\propto&\left[C_{11,1}^{(0)*}\left(C_{11,1}^{(0)}C_{12,1}^{(1)}+C_{11,1}^{(0)}C_{12,1}^{(1)*}-C_{12,1}^{(1)*}C_{12,1}^{(1)} -
      C_{12,2}^{(1)} - C_{12,2}^{(1)*}\right)\right.
    \nonumber \\
    &+& \left.C_{11,2}^{(0)*}\left(C_{12,1}^{(1)} + C_{12,1}^{(1)}
      \right)\right] {\rm Tr}\left(\delta\p^2\F\delta\p\F^3 -
      \delta\p^2\F^3\delta\p\F\right)
      \nonumber \\
      &\propto& {\rm Tr}\left(\F[\F, \delta\p]^3\right) =
      {\rm Tr}\left(\F\right)\det\left([\F,\delta\p]\right).
\end{eqnarray}
where the proportionality constant is just the flavor diagonal product
of Green's functions.
Similar contributions arise from the other terms in \eqref{eq:32} and
all in all we obtain the expected result $\Delta_{\rm CP} \propto
{\rm Tr}(\F)\det([\F, \delta\p])$.

\section{Non-Perturbative Calculations of the Reflection and
  Transmission Coefficients}
\label{sec:non-pert-calc-1}

In this part we calculate the reflection and transmission coefficients
of the fermions in the non-perturbative regime (large Yukawa
couplings).  This is possible at the expense of working within a single
generation framework.  Thus CPV is absent and therefore thermal
effects, in particular thermal masses, are neglected for simplicity.
Our rationale here is that our results below, for the reflection
coefficients in the single flavor case, would also hold when promoting
the model to three generations~\cite{NP1}. The main difference
is the appearance of CPV phases in the thermal masses [see {\it e.g.}
eq. (\ref{eq:3})]. These are expected to be only mildly suppressed by
an order one factor [1/8(1/32) for the quark singlets(doublets) in
the 4D SM case] which we shall take into account in our final
estimation for the resultant BAU\footnote{Note that in \cite{NP1}, CP
  violation originated from a different source, namely 
  twisting~\cite{HPSS,GHPSS}, so that the above mild suppression is
  absent and an asymmetry was induced even in the absence of these
  thermal effects.}.  Consequently, below we shall assume the
following dispersion relations
\begin{equation}
  \label{EG}
  E=\sqrt{{m_n}^2+k_3^2},
\end{equation}
where here and below, in order not to confuse with the perturbative
approach, we use $m_n = p_5^{(n)}$.

As before, for $x_3<0$ the system is in
the unbroken phase while for $x_3>x_3^0$ the system is in the broken
phase.  In order to calculate the reflection and transmission
coefficients we solve the Dirac equations. for the two regions separately
and sew the solutions at $x_3=0$.  As mentioned before the spin up and
down states do not mix in the presence of the localizer's VEV and thus
below we only consider the wave function for the spin down states
where similar derivation can be applied to the spin up case.

\subsection{Unbroken Phase}
For $x^3<0$, the relevant positive frequency solutions are described
by plane waves propagating in the positive $x^3$ direction with
momenta $k_3$.  Since we are dealing with $T\sim R^{-1}$ we will
mostly be interested in the dynamic of the lowest KK states. For each
representation this implies a single zero mode which can be either
incoming or outgoing, but not both. For simplicity we only consider
particle with energies less than $2/R$ and therefore in addition
to the zero-mode, a massive Dirac state corresponding to the first KK
level can be on shell (while higher KK modes are Boltzmann suppressed).
Consequently we also consider incoming or outgoing  massive states.

From (\ref{eq:5}-\ref{eq:27},\ref{eq:8}-\ref{eq:4}) we find that the
solution is characterized by a KK number $n$ (considering as an
example an incoming spin-up fermion):
\begin{equation}
  \chi\left(x^3<0,n,k_3\right)= N_\chi^{\rm u}
  e^{i\left(k_3 x^3-Et\right)} \mypmatrix{ \cos\left(n{ x^5\over
        R}\right) \sqrt{E+k_3}\cr \sin\left(n{ x^5\over R}\right)
    \sqrt{E-k_3}}
  \label{DSolNz}
\end{equation}
where $N_\chi^{\rm u}$ is a
normalization constants and
\begin{equation}
 \label{EUB}
 E =\sqrt{m_n^2+k_3^2}\,,\qquad
 m_n^2={n^2\over R^2}
\end{equation}
Similar expression would hold for the outgoing (reflected) particles
which can be obtained from the above via changing $k_3$ to $-k_3$.
As before, we see that the reflected states do not contain a zero
mode for the spin-up fermions.  Similarly, for spin-down fermions
there is no incoming zero mode.

We can now extract the reflection coefficients at $T\sim 1/R$ by
considering an incoming zero mode with $E\gsim 1/R$ (to allow for an
on shell reflected state) which is reflected to a KK tower of massive
states.  Thus in the unbroken phase the wave function of the
fermions is given by
\begin{eqnarray}
 \Psi^{\downarrow}\left(x^3<0\right)= &&e^{-iE_0 t}\left[\sum_n {R_n}
e^{-ik_3^n x^3}
\mypmatrix{ \cos\left({n
x^5\over R}\right) \sqrt{E-k_3^n\over\pi R}\cr
\sin\left({n x^5\over R}\right) \sqrt{E+k_3^n\over\pi R}
}
+I_0 e^{iE x^3}\sqrt{E\over \pi R}\mypmatrix{1\cr0}\right.
\nonumber \\
&&\left.+ I_1 e^{ik_3^1 x^3}
\mypmatrix{ \cos\left({
x^5\over R}\right) \sqrt{E+k_3^1\over\pi R}\cr
\sin\left({ x^5\over R}\right) \sqrt{E-k_3^1\over\pi R}
}\right]
\label{DUB},  
\end{eqnarray}
where above and below we use normalized
wave functions for the part which depends on $x^5$,
we set the normalization constant for the incoming particle to one and
unless otherwise specified the summation of $n$
is implied and taken from one to the highest KK mode.
As mentioned above we only consider incoming modes with $E<2/R$.
Thus one can distinguish between two cases: (i) the incoming mode is a zero mode, which implies $I_0=1$ and
$I_1=0$, or (ii) the incoming mode is a first KK mode which implies setting $I_1=1$ and
$I_0=0$.

\subsection{Broken Phase}

For $x^3>0$ the general solution depends on $f(x^3,x^5)$ and cannot be
found analytically.  Recall that, for simplicity, we adopt the thin
wall approximation (in the $x^3$ direction) and assume a step function
(in the $x^5$ direction) for the localizer, eq.~\eqref{eq:72}.
To simplify our notations below we use:
\begin{equation}
  \label{eq:50}
   \f(x^3, x^5) = \left\{
    \begin{array}{lll}
      0 && x^3\leq 0 \cr
      f && x^3>0
    \end{array}\right..
\end{equation}
Thus, for a given KK mode, the positive frequency solution is given by:
\begin{equation}
  \label{DSolPz} 
  \chi(x^3>0,n,k'_3)=N^{\rm b}_{\chi n}
  e^{i\left(k'_3 x^3-Et\right)} \mypmatrix{u^+_n(x^5)
    \sqrt{E+k'_3}\cr u^-_n(x^5) \sqrt{E-k'_3}}
\end{equation}
where in this case ${k'_3}^2$, the square of the momentum in the $x^3$
direction, can be either negative or positive, corresponding to on
shell and virtual modes respectively.  Solving for the $x^5$ dependent
part, using eqs.  (\ref{eq:5}-\ref{eq:27},\ref{eq:8}-\ref{eq:4}), we
find that there are two types of solutions: an exponential solution 
corresponding to a zero mode,
\begin{equation}
  \label{eq:49}
  u^+_0(x^5)=\sqrt{f\over e^{2\pi
      f R}-1}\,e^{f x^5}\,,\ \ \ u^-_0(x^5)=0\,, \qquad
  E^2={k'_3}^2=k_3^2 
\end{equation}
and an oscillatory solution which corresponds
to higher KK modes,
\begin{eqnarray}
  u_n^+(x^5)&=&\sqrt{n^2\over \pi
    R\left( f^2 R^2+n^2\right)}\,\cdot\,
  \left[\cos\left(k^n_5 x^5\right)+{f\over k^n_5}\sin\left(k^n_5
      x^5\right)\right]\,,\label{evenWF}
  \nonumber \\
  u_n^-(x^5)&=&\sqrt{1\over \pi R}\,\cdot\, \sin\left(k^n_5
    x^5\right)\,,\qquad \qquad E^2={k'_3}^2+
  {k^n_5}^2={k'_3}^2+f^2+{n^2\over R^2} \,.
  \label{oddWF}
\end{eqnarray}
Hence in the broken phase the wave function of an outgoing
fermion with positive spin is given by
\begin{eqnarray}
  \begin{split}
    \chi\left(x^3>0\right)=e^{-iE t}&\Bigg\{ \sum_n {T_n}
      e^{-i{k'_3}^n x^3}
      \mypmatrix{\left[ \cos\left({n
              x^5\over R}\right)+{f\over k^n_5}\sin\left({n
              x^5\over R}\right) \right]\sqrt{E+{k'_3}^n\over \pi R}\sqrt{n^2\over
          n^2+f^2 R^2}\cr
        \sin\left({n x^5\over R}\right) \sqrt{E-{k'_3}^n\over \pi R}}
      \\
      &+ 
      {T_0} e^{iE x^3+f x^5} \sqrt{2E f\over e^{2\pi f
          R}-1}\mypmatrix{1\cr0}\Bigg\},
  \end{split}
    \label{DB},
\end{eqnarray}
where
\begin{equation}
  \label{kp3n}
  {k'_3}^n=\sqrt{E^2-m_n^2}=\sqrt{E^2-f^2-{n^2\over R^2}}.
\end{equation}
It is important to note that since in our framework $fR$ is large
(to produce significant localization), $fR\sim 5-10\,,$
the non-zero modes in the broken phase are very heavy.
Thus only the zero mode is produced on-shell and be transmitted
through from the unbroken phase.

\subsection{Computation of the Reflection and Transmission
  Coefficients}
We can equate the expressions for the wave functions
in the broken and unbroken phase, and extract the reflection and
transmission coefficients.  Below we distinguish between the two
possibilities for an incoming state, namely between an incoming zero
mode and an incoming first KK mode.  As a consistency check we also
verify below that our solutions make sense in the $f\to0$ limit 
which can be computed analytically. 

\subsubsection{Reflection Coefficients for an Incoming
  Zero Mode} The reflection and transmission coefficients are found by
requiring that the wave function is continuous at $x^3=0$. This is
done by comparing the expressions in \eqref{DUB} and \eqref{DB}.  We
note that the functional dependence on $x^5$ for the lower component
(negative helicity) in both cases is equal up to a kinematic factor.
This implies the following relation between the coefficients:
\begin{equation}
  \label{compare}
  {{T_n^2}\over {R_n^2}}={E+k_3^n\over
  E-{k'_3}^n}= {E+\sqrt{E^2-{n^2\over R^2}}\over
  E-\sqrt{E_0^2-f^2-{n^2\over R^2}}}.
\end{equation}
Substituting the
above relation into the upper component of (\ref{DUB},\ref{DB}) we get
the following equation for the reflection coefficients (after
multiplying both sides by $\sqrt {\pi R \over E} $):
\begin{eqnarray}
  \label{Rrelation}
   1+ \sum_n
  {R_n} e_n \cos\left({n w}\right)
  &=&\sqrt{2\pi Y \over e^{2\pi Y}-1} {T_0} e^{wY}+
  \nonumber\\
  &&\sum_n {R_n}e_n A_n \left[\cos\left({nw}\right)+{Y\over n}
    \sin\left({nw}\right)\right]
\end{eqnarray}
where
\begin{eqnarray}
 \label{Def}
 w &\equiv& x^5/R\,,\qquad X \equiv  E R\,,\qquad Y \equiv f R
 \nonumber\\ \nonumber \\
 A_n&\equiv& \sqrt{n^2\over Y^2+n^2} \,\cdot\,\sqrt{E+k_3^n\over E-k_3^n}
 \,\cdot\,\sqrt{E+{k'_3}^n\over E-{k'_3}^n}\,,
 \qquad e_n\equiv \sqrt{1-{k_3^n \over E}}\, \qquad n\neq0\,.
\end{eqnarray}
In order to find the values of the various coefficients
we expand all functions in the cosine basis,
\begin{eqnarray}
  \label{expan}
  e^{wY}&=&{1\over\pi}\,\cdot\,{e^{Y\pi}-1\over Y}+ {2Y\over\pi} \sum_{n} { (-1)^n
    e^{Y\pi}-1\over Y^2+n^2} \cos\left({nw}\right),
  \nonumber\\
  \sin\left({nw}\right)&=&{1\over\pi}\,\cdot\,{1-(-1)^{n}\over n}
  +{ 2n\over \pi} \sum_m {1-(-1)^{n+m}\over
    n^2-m^2}\cos\left({mw}\right)\,, \ \ \ m\neq n
\end{eqnarray}
where in the second line if $m=n$ the coefficient is zero.  Substitute
(\ref{expan}) into (\ref{Rrelation}) and collecting terms we get:
\begin{eqnarray}
 \label{Rrelation1}
 &&1-\sqrt{2\over \pi Y}\,\cdot\,{T_0}\,\cdot\,\sqrt{e^{Y\pi}-1\over
   e^{Y\pi}+1}- {Y\over\pi}\,\cdot\,\sum_n {R_n}e_n
 A_n\,\cdot\,{1-(-1)^{n}\over n^2}
 \nonumber\\
 &+& \sum_{nm}\left\{ {R_n}e_n\left[ \delta_{nm}-A_n\,\cdot\,
     \left(\delta_{nm}+ {2Y\over \pi}\cdot{1-(-1)^{n+m}\over
         n^2-m^2}\left(1-\delta_{nm}\right)\right)\right]\right.\nonumber\\
 &-&\left.  N_{\chi 0}^{\rm b}\,\sqrt{2Y\over\pi}\, \cdot\,{2Y\over
     Y^2+n^2}\,\cdot\,{ (-1)^n e^{Y\pi}-1\over \sqrt{e^{2Y\pi}-1}
   }\,\delta_{nm} \right\}\cos\left({m w}\right)=0
\end{eqnarray}
where summation over $n,m$ begins at one.

We can use the relation for the zeroth element to express
${T_0}$ as a function all the other unknowns,
\begin{eqnarray}
 \label{RT0rel}
  {T_0}=\sqrt{{Y\pi\over2}\,\cdot\,{ e^{Y\pi}+1\over e^{Y\pi}-1}}
\left[1-{Y\over\pi}\sum_n {R_n}e_n A_n\,\cdot\,{1-(-1)^{n}\over n^2}\right].
\end{eqnarray}
Substituting \eqref{RT0rel} back into \eqref{Rrelation1} we get,
\begin{eqnarray}
  \label{Rrelation2}
  &&\sum_{nm}{R_n}e_n\left\{ \delta_{nm}-A_n\, \left[\delta_{nm}+
    {2Y\over \pi}\cdot{1-(-1)^{n+m}\over
      n^2-m^2}\left(1-\delta_{nm}\right)\right.\right.\nonumber\\
&-&\left.\left.  {1-(-1)^{n}\over\pi n^2} \, \cdot\,{2Y^3\over
      Y^2+m^2}\,\cdot\,{ (-1)^m e^{Y\pi}-1\over e^{Y\pi}-1}
  \right]\right\}\cos\left({m w}\right)\nonumber\\
&=& {2Y^2\over e^{Y\pi}-1}\,\sum_m\,{ (-1)^m e^{Y\pi}-1\over Y^2+m^2}
\,\cdot\,\cos\left({m w}\right).
\end{eqnarray}
Let us write the above relation in a matrix form, describing the set of equations for the various
Fourier coefficients
\begin{eqnarray}
 \label{Mat}
  M^{R_n}_{mn}{R_n}=V_n,
\end{eqnarray}
where $M^{R_n}_{mn}$ and $V_n$ can be
directly extracted from (\ref{Rrelation2}).  Inverting $M$ yields the
solution for the reflection coefficients
\begin{eqnarray}
 \label{SolMat}
  {R_n}=\left(M^{R_n}_{mn}\right)^{-1}\, V_n,
\end{eqnarray}
This of course is done numerically.

\subsubsection{Reflection Coefficients for an Incoming
First KK Mode with Spin Up}
As before, we write the relation between the reflection and
transmission coefficients, this time for higher KK modes, 
\begin{eqnarray}
  \label{compare1}
  {T_n}=\left\{ \begin{array}{lll}
      {R_n}\sqrt{E_0+k_3^n\over
        E_0-{k'_3}^{n}}&\qquad n\neq1 \\
      {R_n}\sqrt{E_0+k_3^n\over
        E_0-{k'_3}^{n}}+\sqrt{E_0-k_3^n\over
        E_0-{k'_3}^{n}}&\qquad n=1\end{array}
  \right..
\end{eqnarray}
Again we substitute the above to get,
\begin{eqnarray}
  \label{Rrelation11}
  && \sqrt{1+{k_3^1 \over E}} \cos w +\sum_{n} {R_n} e_n
  \cos\left({n w}\right)
  =\sqrt{2\pi Y \over e^{2\pi Y}-1} {T_0} e^{wY}+
  \nonumber\\
  &&\sum_n\e_n\left( {R_n} A_n +B_n\delta_{1n}\right)
  \left[\cos\left({nw}\right)+{Y\over n} \sin\left({nw}\right)\right]
\end{eqnarray}
where
\begin{eqnarray}
  \label{Def1}
  B_n&\equiv& \sqrt{n^2\over Y^2+n^2} 
  \,\cdot\,\sqrt{E+{k'_3}^n\over E-{k'_3}^n}
\end{eqnarray}

Expanding and collecting terms, using \eqref{expan}, one obtains,
\begin{eqnarray}
  \label{Rrelation12}
  &&-\sqrt{2\over \pi Y}\,\cdot\,{T_0}\,\cdot\,\sqrt{e^{Y\pi}-1\over e^{Y\pi}+1}-
  {Y\over\pi}\,\cdot\,\sum_n e_n {R_n} A_n \,\cdot\,{1-(-1)^{n}\over
    n^2}-{2Y\over\pi} e_1 B_1
  \nonumber\\
  &+&
  \sum_{nm}\left\{ {R_n}e_n\left[
      \delta_{nm}-A_n\,\cdot\, \left(\delta_{nm}+
        {2Y\over \pi}\cdot{1-(-1)^{n+m}\over
          n^2-m^2}\left(1-\delta_{nm}\right)\right)\right]\right.
  \nonumber\\
  &-&\left.
    {T_0}\,\sqrt{2Y\over\pi}\, \cdot\,{2Y\over Y^2+n^2}\,\cdot\,{ (-1)^n e^{Y\pi}-1\over
      \sqrt{e^{2Y\pi}-1} }\,\delta_{nm}
  \right\}\cos\left({m w}\right)
  \nonumber\\
  &=&\left(e_1 B_1- \sqrt{1+{k_3^1 \over E}}\right) \,\cdot\,\cos w+ e_1 B_1\sum_{m=2}
  {2Y\over \pi}\cdot{1+(-1)^{m}\over
    1-m^2}\,\cdot\,\cos\left({m w}\right)
\end{eqnarray}
which leads to,
\begin{eqnarray}
  \label{RT0rel11}
  {T_0}=-\sqrt{{Y\pi\over2}\,\cdot\,{ e^{Y\pi}+1\over e^{Y\pi}-1}}
\left[{2Y\over\pi} e_1 B_1+{Y\over\pi}\sum_n {R_n}e_n A_n\,\cdot\,{1-(-1)^{n}\over n^2}\right].
\end{eqnarray}
Substituting \eqref{RT0rel11} in \eqref{Rrelation12} we get,
\begin{eqnarray}
  \label{Rrelation21}
  &&\sum_{nm}{R_n}e_n\left\{
    \delta_{nm}-A_n\, \left[\delta_{nm}+
      {2Y\over \pi}\cdot{1-(-1)^{n+m}\over
        n^2-m^2}\left(1-\delta_{nm}\right)\right.\right.\nonumber\\
  &-&\left.\left.
      {1-(-1)^{n}\over\pi n^2}
      \, \cdot\,{2Y^3\over Y^2+m^2}\,\cdot\,{ (-1)^m e^{Y\pi}-1\over 
        e^{Y\pi}-1}
    \right]\right\}\cos\left({m u}\right)
  =\left(e_1 B_1- \sqrt{1+{k_3^1 \over E}}\right) \,\cos u\nonumber\\
  &+& e_1 B_1\,{2Y\over \pi}\,\sum_{m}\left[
    {1+(-1)^{m}\over
      1-m^2}\,\left(1-\delta_{1m}\right)-{2Y^2\over Y^2+m^2}\,\cdot\,{ (-1)^m e^{Y\pi}-1\over 
      e^{Y\pi}-1}\right]\,\cos\left({m u}\right),
\end{eqnarray}
which may again be used to extract the reflection coefficients.

\subsubsection{Reflection Coefficients for an Incoming
First KK Mode with Spin Down}
This time,
\begin{eqnarray}
  {T_n}=\left\{ \begin{array}{lll}
      {R_n}\sqrt{E_0-k_3^n\over
        E_0+{k'_3}^{n}}\,,&\qquad n\neq1\,,\cr
      {R_n}\sqrt{E_0-k_3^n\over
        E_0+{k'_3}^{n}}+\sqrt{E_0+k_3^n\over
        E_0+{k'_3}^{n}}\,,&\qquad n=1\,;\end{array}
  \right.
  \label{compare11}
\end{eqnarray}
which, after substitution, gives, 
\begin{eqnarray}
 \label{Rrelation112}
  &&{R_0}+\sqrt{1-{k_3^1 \over E}} \cos w +\sum_{n} N_{\rm Rn}^{\rm u} e'_n
\cos\left({n w}\right)
=\nonumber\\
&&\sum_n\e'_n\left( {R_n} A'_n +B'_n\delta_{1n}\right)
\left[\cos\left({nw}\right)+{Y\over n} \sin\left({nw}\right)\right],
\end{eqnarray}
with
\begin{eqnarray}
  \label{Def12}
  A'_n&\equiv& \sqrt{n^2\over Y^2+n^2} \,\cdot\,\sqrt{E-k_3^n\over E+k_3^n}
\,\cdot\,\sqrt{E-{k'_3}^n\over E+{k'_3}^n}\,,
\qquad e'_n\equiv \sqrt{1+{k_3^n \over E}}\,,\nonumber\\
B'_n&\equiv& \sqrt{n^2\over Y^2+n^2} 
\,\cdot\,\sqrt{E-{k'_3}^n\over E+{k'_3}^n}.
\end{eqnarray}
Finally, repeating the same procedure as before, one gets,
\begin{eqnarray}
  \label{Rrelation121}
  &&{R_0}-
{Y\over\pi}\,\cdot\,\sum_n e'_n N_{\rm Rn}^{\rm u} A'_n \,\cdot\,{1-(-1)^{n}\over
 n^2}-{2Y\over\pi} e'_1 B'_1
\nonumber\\
&+&
 \sum_{nm}\left\{ N_{\rm Rn}^{\rm u}e'_n\left[
\delta_{nm}-A'_n\,\cdot\, \left(\delta_{nm}+
{2Y\over \pi}\cdot{1-(-1)^{n+m}\over
  n^2-m^2}\left(1-\delta_{nm}\right)\right)\right]\right\}\cos\left({m w}\right)\nonumber\\
&=&\left(e'_1 B'_1- \sqrt{1-{k_3^1 \over E}}\right) \,\cdot\,\cos w+ e'_1 B'_1\sum_{m=2}
{2Y\over \pi}\cdot{1+(-1)^{m}\over
  1-m^2}\,\cdot\,\cos\left({m w}\right).
\end{eqnarray}

\subsubsection{Check $Y\to0$}
In order to obtain some intuition and check our results, let us check
the above in the small energy, small barrier limit (see \eqref{Def}),
\begin{equation}
 \label{Y0}
  Y\to0.
\end{equation}
In that limit, to leading order, relation \eqref{Rrelation2} is
simplified
\begin{eqnarray}
  \label{Rrel2app}
  && \sum_{nm}
  {R_n}e_n\left(A_n-1\right)\delta_{mn}\cos\left({m w}\right)
  = {2Y\over\pi}\sum_m{ 1-(-1)^m \over m^2}\cos\left({m w}\right),
\end{eqnarray}
and we find the following expressions for $A_n$ \eqref{Def},
\begin{eqnarray}
  \label{eq:86}
  A_n= -1+2{X\over n}\sqrt{{X^2\over n^2}-1}+2{X^2\over n^2}
+{\cal O}\left(Y^2\right)\,.
\end{eqnarray}

Under the above approximation the explicit matrix $M_{mn}$ (\ref{Mat})
can easily be inverted and we find the following approximated
solution for the reflection coefficients:\footnote{The reason that the
  even coefficients vanish is probably related to fact that the
  potential is symmetric with respect to the reflection around the point
  $w=\pi/2$.  Thus to first order only transitions between states that
  respect this symmetry are allowed.}
\begin{eqnarray}
 \label{Rnapsol}
 {R_{2n-1}}=
 {4Y\over\pi \left(A_{2n-1}-1\right) e_{2n-1} \left(2n-1\right)^2 }\,,
 \qquad {R_{2n}}=0 .
\end{eqnarray}
Furthermore, using \eqref{RT0rel} we also have the transmission coefficient,
\begin{eqnarray}
 \label{RT0apSol}
  {T_0}=\left[1+{\cal
     O}\left(Y^2\right)\right] \left[1-{8Y^2\over\pi^2}
   \sum_{2n-1}\,\cdot\,{ A_{2n-1}\over
     \left(A_{2n-1}-1\right)\left(2n-1\right)^4}\right].
\end{eqnarray}

The above simplifications and the generalizations to the reflections
of other modes, allow us to apply several checks for our numerical results.
\begin{itemize}
\item Vanishing of odd reflection coefficients for small $Y$.  Verified numerically.
\item When $X\leq1$ we expect $N_{\chi 0}^{\rm b}$ to be one.  Verified numerically. 
\item For $2\geq X\geq1$ and $Y\ll1$
  we expect $N_{\chi 0}^{\rm b}$ to deviate from unity only at the
  order of $Y^2$.  The reflection coefficient of the first KK mode, is
  expected to deviate at order $Y$. Verified numerically.
\end{itemize}
Another way to verify that our numerical results is correct is to check that it
is satisfying the unitarity condition as we discuss next.
 
\section{Unitarity Conditions}
\label{apUnit}
We now wish to establish the unitarity conditions for our five
dimensional scattering process.   To obtain intuition, we consider the relation between the reflection and transmission
coefficient in the following two cases: (i) a non-relativistic scalar - one particle quantum mechanics described via
Schroedinger equation and (ii) relativistic spin half fermion - one
particle quantum mechanics described via Dirac equation.

\subsection{Schroedinger Equation}
We wish to calculate the reflection of transmission
coefficients of a particle in a step function potential:
\beq
V(x)=V\theta(x^3)\,.
\eeq
The Schroedinger equation reads:
\beq
\left(-\nabla^2+V(x)\right) \psi=E\psi
\,.
\eeq
For an incoming particle with momentum $\vec k=(0,0,k)$
the solution is given by
\begin{eqnarray}
  \label{ansS}
  \psi(x^3)=\theta(-x^3)\left[e^{ik x^3}+re^{-ik x^3}\right]
  +\theta(x^3) t e^{ik' x^3}\,\qquad k'^2=E-V.
\end{eqnarray}
Continuity of the wave function and its derivative requires,
\begin{eqnarray}
  \label{solS}
  r={1-R_k\over 1+R_k}\,,\qquad t={2\over 1+R_k}\,,\qquad R_k={k'\over k}.
\end{eqnarray}

To obtain the unitarity condition, we consider the Schroedinger
conserved current equation (for static density all over space) which
is given by
\begin{eqnarray}
  \label{eq:39}
  \vec\nabla\cdot(\psi^*\vec\nabla\psi-\psi\vec\nabla\psi^*)=0.
\end{eqnarray}
Integrating the above from $-\infty$ to $\infty$ and omitting the
interference term we find
\begin{eqnarray}
  \label{eq:48}
  \left(\psi^*\vec\nabla\psi-\psi\vec\nabla\psi^*\right)|^\infty_{-\infty}=0,
\end{eqnarray}
which implies for \eqref{ansS} the relation:
\begin{eqnarray}
  \label{eq:51}
  |r|^2+R_k|t|^2=1,
\end{eqnarray}
and clearly (\ref{solS}) is consistent with the above.

\subsection{Dirac Equation - 4D Case}
Let us first consider a reflection problem from a one dimensional wall
embedded in four dimensions.
The Dirac equation reads,
\begin{eqnarray}
  \label{eq:52}
  \gamma^0\left[\nabla\cdot\vec\gamma+V(x)\right] \psi=E\psi.
\end{eqnarray}
For a massless incoming spin down particle with momentum $k^\mu=(k,0,0,k)$ the solution is
(in the chiral basis),
\begin{eqnarray}
 \label{ansD}
  \psi(x^3)&=&
\theta(-x^3)\left[e^{ik x^3}\mypmatrix{0\cr\sqrt{2E}\cr0\cr 0}+re^{-ik x^3}\mypmatrix{0\cr0\cr0\cr\sqrt{2E}}\right]
+\theta(x^3) t e^{ik' x^3}\mypmatrix{0\cr\sqrt{E+k'}\cr0\cr
  \sqrt{E-k'}}\,\nonumber\\
k'^2&=&E^2-V^2.
\end{eqnarray}
Note that in this case (which is different from the 5D case) the
transmitted particle is massive while the reflected one is massless.
From continuouity of the wave function one finds:
\begin{eqnarray}
  \label{solD}
  r=\sqrt{1-R_k\over 1+R_k}\,,\qquad t=\sqrt{2\over 1+R_k}\,,\qquad
  R_k={k'\over k}.
\end{eqnarray}

Again, unitarity condition is found through the Dirac conserved
current equation,
\begin{eqnarray}
  \label{eq:55}
  \vec\nabla\cdot\left(\bar\psi\vec\gamma\psi\right)=0,
\end{eqnarray}
which gives
\begin{eqnarray}
  \label{eq:56}
  \left(\bar\psi\vec\gamma\psi\right)|^\infty_{-\infty}=0.
\end{eqnarray}
We thus obtain the condition,
\begin{eqnarray}
  \label{eq:57}
  |r|^2+R_k|t|^2=1.
\end{eqnarray}

\subsection{Dirac Equation - 5D Case}
We now move on to discuss the 5D problem. For finite
temperature of order $(1/R)$ we assume it is enough to consider only the
zero modes and the first KK state.

\subsubsection{Incoming Spin-Down Massless Particle}
Thus we will assume that the incoming particle has energy slightly above $(1/R)$
and that it is reflected into an on shell first excited KK state while
the on shell transmitted particle is massless.   
The 5D Dirac equation reads
\begin{eqnarray}
  \label{eq:58}
  \gamma^0\left[\nabla\cdot\vec\gamma+V(x)+\gamma_5\partial_5\right] \psi=E\psi.
\end{eqnarray}
For a massless incoming particle with momentum $k^\mu=(k,0,0,k)$
the solution is (again in the chiral basis)
\begin{eqnarray}
 \label{ans5D}
 \psi(x^3)&=&
 \theta(-x^3)\left[e^{ik x^3}\mypmatrix{0\cr\sqrt{2E}\cr0\cr 0}+re^{-ik'
     x^3}
   \mypmatrix{0\cr\sqrt{E-k'}\cr0\cr
     \sqrt{E+k'}}\right]
 +\theta(x^3) t e^{ik x^3}\mypmatrix{0\cr\sqrt{2E}\cr0\cr0}\,\nonumber\\
 k'^2&=&E^2-{1\over R^2}.
\end{eqnarray}
Since the extra dimension is compact, there is no current flowing in
this direction.  The conserved current is therefore
\begin{eqnarray}
  \label{eq:53}
  \vec\nabla\cdot\left(\bar\psi^*\vec\gamma\psi\right)=0,
\end{eqnarray}
which implies,
\begin{equation}
 \label{U5D}
  R_k|r|^2+|t|^2=1.
\end{equation}

\subsubsection{Incoming Spin-Down First KK Mode}
For a massive incoming particle with momentum $k^\mu=(k,0,0,k')$ the
solution this time is
\begin{eqnarray}
 \label{ans5D1}
 \psi(x^3)&=&
 \theta(-x^3)\left[e^{ik'
     x^3}
   \mypmatrix{0\cr\sqrt{E+k'}\cr0\cr
     \sqrt{E-k'}}+re^{-ik'
     x^3}
   \mypmatrix{0\cr\sqrt{E-k'}\cr0\cr
     \sqrt{E+k'}}\right]
 +\theta(x^3) t e^{ik x^3}\mypmatrix{0\cr\sqrt{2E}\cr0\cr0}\,
 \nonumber\\
 k'^2&=&E^2-{1\over R^2}.
\end{eqnarray}
As before, one obtains the relation
\begin{equation}
  \label{U5D1}
  |r|^2+R_k^{-1}|t|^2=1.
\end{equation}

\subsubsection{Incoming Spin-Up First KK Mode}
Here the solution is
\begin{eqnarray}
  \label{ans5D12}
  \psi(x^3)&=&
  \theta(-x^3)\left[e^{ik'
      x^3}
    \mypmatrix{\sqrt{E-k'}\cr0\cr
      \sqrt{E+k'}\cr0}+r_1 e^{-ik'
      x^3}
    \mypmatrix{\sqrt{E+k'}\cr0\cr
      \sqrt{E-k'}\cr0}\right]
  +\theta(-x^3) r_0 e^{-ik x^3}\mypmatrix{\sqrt{2E}\cr0\cr0\cr0}\,\nonumber\\
  k'^2&=&E^2-{1\over R^2},
\end{eqnarray}
and therefore the unitarity condition
\begin{eqnarray}
 \label{U5D12}
  |r_1|^2+R_k^{-1}|r_0|^2=1.
\end{eqnarray}

\end{document}